\documentstyle[12pt,aps, float,epsfig]{revtex}

\newcommand{\beq}{\begin{equation}}
\newcommand{\eeq}{\end{equation}}

\newcommand{\bfr}{{\bf r}}

\newcommand{\zd}{\overline{z}}

\newcommand{\nablad}{ \overline{\nabla}}
\newcommand{\derz}{\partial_{i}}
\newcommand{\derzj}{\partial_{j}}
\newcommand{\derzd}{\bar \partial_{i}}
\newcommand{\derzdj}{\bar \partial_{j}}

\newcommand{\be}{\begin{equation}}
\newcommand{\ee}{\end{equation}}
\newcommand{\bea}{\begin{eqnarray}}
\newcommand{\eea}{\end{eqnarray}}
\newcommand{\e}{\varepsilon}

\newcommand{\ba}{\begin{array}}   %
\newcommand{\ea}{\end{array}}     %
\newcommand{ \jd}{\mbox{ \tiny\put( 0,0){$\Box$}\put( 1.1,0){$\Box$}}
\hspace{ 0.24cm} }
\newcommand{ \juu}{\mbox{ \tiny\put( 0,0){$\Box$}\put( 0,1.1){$\Box$}}
\hspace{ 0.13cm} }
\newcommand{ \jt}{\mbox{ \tiny\put( 0,0){$\Box$}\put( 1.1,0){$\Box$}
\put( 2.2,0){$\Box$}} \hspace{ 0.35cm} }
\newcommand{ \jdu}{\mbox{ \tiny\put( 0,1.1){$\Box$}\put( 1.1,1.1){$\Box$}
\put( 0,0){$\Box$}} \hspace{ 0.24cm} }
\newcommand{ \juuu}{\mbox{ \tiny\put( 0,2.2){$\Box$}\put( 0,1.1){$\Box$}
\put( 0,0){$\Box$}} \hspace{ 0.13cm} }
\newcommand{ \jq}{\mbox{ \tiny\put( 0,0){$\Box$}\put( 1.1,0){$\Box$}
\put( 2.2,0){$\Box$}\put( 3.3,0){$\Box$}} \hspace{ 0.46cm} }
\newcommand{ \jtu}{\mbox{ \tiny\put( 0,1.1){$\Box$}\put( 1.1,1.1){$\Box$}
\put( 2.2,1.1){$\Box$}\put( 0,0){$\Box$}} \hspace{ 0.35cm} }
\newcommand{ \jdd}{\mbox{ \tiny\put( 0,1.1){$\Box$}\put( 1.1,1.1){$\Box$}
\put( 0,0){$\Box$}\put( 1.1,0){$\Box$}} \hspace{ 0.24cm} }
\newcommand{ \jduu}{\mbox{ \tiny\put( 0,2.2){$\Box$}\put( 1.1,2.2){$\Box$}
\put( 0,1.1){$\Box$}\put( 0,0){$\Box$}} \hspace{ 0.24cm} }
\newcommand{ \juuuu}{\mbox{ \tiny\put( 0,3.3){$\Box$}\put( 0,2.2){$\Box$}
\put( 0,1.1){$\Box$}\put( 0,0){$\Box$}} \hspace{ 0.13cm} }

\begin{document}
\baselineskip=17pt

\gdef\journal#1, #2, #3, 1#4#5#6{{#1~}{\bf #2}, #3 (1#4#5#6)}
\gdef\ibid#1, #2, 1#3#4#5{{\bf #1} (1#3#4#5) #2}

\begin{flushright}
January 1999 \\
hep-th/9902028
\end{flushright}
\sloppy
\vskip 0.8cm

\centerline {\Large Non-Abelian Chern-Simons Particles in an External \rm}
\vskip 0.1cm
\centerline {\Large  Magnetic Field }
\vskip 1cm
\centerline {\large  Serguei B. Isakov, Gustavo Lozano, St\'ephane Ouvry }
\vskip 0.2cm
\centerline {Laboratoire de Physique Th\'eorique et Mod\`eles 
Statistiques\footnote{\it Unit\a'e de
Recherche de l' Universit\a'e Paris 11 associ\a'ee au CNRS, UMR 8626}}
\centerline {B\^at. 100, Universit\'e Paris-Sud, 91405 Orsay, France}

\vskip 1cm
\centerline{\large Abstract}
\vskip 0.2cm

The quantum mechanics and thermodynamics of SU(2)
non-Abelian Chern-Simons particles (non-Abelian anyons)
in an external magnetic field are addressed.
We derive the $N$-body Hamiltonian in the (anti-)holomorphic gauge 
when the Hilbert space is projected onto the lowest Landau level of the
magnetic field.
In the presence of an additional harmonic potential, 
the $N$-body spectrum  depends linearly on the coupling 
(statistics) parameter. We calculate the second virial coefficient
and find that in the strong magnetic field limit it develops
a step-wise behavior as a function of the statistics parameter, 
in contrast to the linear dependence in the case of Abelian anyons.
For small enough values of the statistics parameter
we relate the  $N$-body partition functions    
in the lowest Landau level to those of SU(2) bosons and find that
the cluster (and virial) coefficients dependence on the statistics parameter
cancels.


\newpage
\section[]{Introduction}

Identical particles with statistics continuously interpolating between
Bose-Einstein and Fermi-Dirac statistics exist in two and one dimensions.
Explicit implementations of these ideas can be found in 
the two-dimensional anyon model 
\cite{Anyon} and in the one-dimensional
Calogero-Sutherland  models \cite{Calogero-Sutherland} 
where a different approach to statistics proposed 
by Haldane based on a generalized Pauli exclusion principle \cite{Haldane} is
realized.

Contrary to the $N$-body Calogero model, which is solvable, the $N$-anyon
spectrum is still unknown. However, a simplification arises when considering the anyon model in
the background of an external magnetic field. 
By projecting this model onto the lowest Landau level, 
a procedure which is justified in the strong field-low temperature
limit, a complete eigenstate basis, which continuously interpolates
between the bosonic and the fermionic  basis, can be found
in the screening regime where the flux $\phi$
carried by the anyons  is antiparallel to the 
external magnetic field $B$.  More precisely, 
when the statistics parameter $\alpha=\phi/\phi_o$, 
which varies from  $\alpha=0$ to
$\alpha=\pm 1$, is such that 
$\alpha\in [-1,0]$ if $eB>0$, or
equivalently $\alpha\in [0,1]$ if $eB<0$.

In this situation, the statistical mechanical properties  of 
the anyon gas can be studied in a complete and explicit way \cite{Notre}
and they turn out to be quite similar to those of the Calogero model. 
In the thermodynamic limit both models are  microscopical realizations of
Haldane statistics
\cite{Haldane,ES,ES-integrable}. 
Recently, various conformal field  theories have also been
shown to implement exclusion statistics \cite{SchoutensES-CFT}.

It is well known that anyons can be thought of as
bosons or fermions coupled to Abelian Chern-Simons gauge fields. 
An interesting generalization
occurs when the gauge fields take values in 
a non-Abelian group and the particles carry
internal degrees of freedom associated with a representation
of this group. 
These models describing
non-Abelian Chern-Simons particles
 have already been considered
in several contexts. 
Verlinde \cite{Verlinde} argued that  they 
provide an explicit realization of non-Abelian {\it braiding} statistics, 
i.e. statistics corresponding to non-Abelian irreducible representations of 
the braid group \cite{nonabelian_braiding}.
Field theoretical implementations  
of such models were recently proposed \cite{GLCSnonAbelian} as  
Ginzburg-Landau Chern-Simons theories 
for Pfaffian (non Abelian) quantum Hall states.
They generalize Abelian Chern-Simons 
field theories \cite{AbelianGLCS}
for Laughlin (Abelian) quantum Hall states.

Whether models of non-Abelian Chern-Simons particles
restricted to the lowest Landau level can be solved exactly (a natural 
possibility since they become effectively one-dimensional),
and whether their thermodynamics  yield a  realization of
exclusion statistics  different from that corresponding to Abelian anyons,
are questions of interest.
In this paper, we address these questions by studying the simplest case of
a non-Abelian symmetry, namely SU(2) Chern-Simons particles
in the fundamental representation. The paper is organized
as follows.

In Section \ref{Verlinde} the Verlinde model \cite{Verlinde},
a generalization of
the Aharonov-Bohm Hamiltonian in the (anti-)holomorphic gauge
to the non-Abelian case, is introduced. We include contact  $\delta^2(z_i-z_j) $
interactions to enforce that the wave functions are regular
when  particles approach each other, and,
 in addition, a confining harmonic
potential to lift the lowest Landau level degeneracy.
We redefine the wave function essentially
as $\psi = U \psi '$, where $U$ 
(the non-Abelian generalization of  $\prod_{i<j} (z_i -z_j)^{\alpha}$
for Abelian anyons with statistics parameter $\alpha$) is a solution of the 
 Knizhnik-Zamolodchikov equation. 
 Our  main result is the Hamiltonian acting on 
 $\psi '$, which takes particularly simple form when $\psi '$ is 
(anti-)holomorphic, that is 
when the Hilbert space is restricted to
the lowest Landau level of the external magnetic field. 
In the latter case we find a linear dependence of 
the $N$-body energy spectrum on the statistics parameter, generalizing 
the known results for Abelian anyons.

In Section \ref{revisited} we revisit both the Abelian and non-Abelian 
anyon models to present another derivation of the same Hamiltonian 
acting on $\psi '$  
as in Section \ref{Verlinde}, but now  starting from  free Pauli Hamiltonians. 
 In this approach, contact interactions do not need to be introduced, 
 instead the spin coupling plays a crucial role.
 We also show that 
in the redefinition of the wave function the external magnetic field 
and the magnetic fluxes carried by  particles can be treated on the 
same footing. The latter point of view is more adapted for models that use 
mean field approximations \cite{Desbois}.

In Section \ref{2-body} we discuss the 2-body problem 
for an arbitrary strength of the external magnetic field. 
The 2-body problem decomposes into two Abelian anyon problems,  
which allows one to calculate the second virial coefficient exactly.
In the vanishing magnetic field limit
our result reduces to that  found  previously by Hagen \cite{Hagen}
in his comment on the paper by Lee \cite{LeePRL}.
In the strong magnetic field limit we find that 
the interpolation from Bose to Fermi statistics is
 different from the Abelian one.
In the latter case \cite{Notre}, which yields Haldane's generalization of the
Pauli exclusion principle, the second virial coefficient 
depends linearly on the statistics parameter. 
Contrary to that, we find that 
in the non-Abelian case the second virial coefficient 
develops a step-wise behavior as a function of the statistics parameter.

In Section \ref{N-body} we address the $N$-body problem in the lowest 
Landau level.
We show how to calculate the partition function taking  into account
properly degeneracies associated with internal isospin degrees of freedom.
The $N$-body energy spectrum
for a given isospin is the $N$-body  SU(2) bosonic spectrum plus a
 a term linear in the statistics parameter.
We relate the $N$-body bosonic partition functions $Z_{N,I}$
with  total isospin $I$ 
to the partition functions associated with Young diagrams.
The latter  partition functions,
which turn out to be generalizations of the Schur functions,
 are introduced in subsection  \ref{Z_Y}.
We propose systematic rules to calculate them.
Collecting all these results, we can
calculate the cluster and virial coefficients one by one.
For small enough values of the statistics parameter
we find, somewhat surprisingly, a
 cancellation of the dependence on the statistics
parameter, and thus  the same cluster and virial coefficients as
for SU(2) bosons.
The $N$-body lowest Landau level thermodynamics of 
non-Abelian Chern-Simons particles  
in the entire interval of definition of the statistics parameter is 
yet an open question. 

We conclude in Section \ref{conclusion}, and comment also
on generalizations to other symmetry groups and relations to one-dimensional
integrable models with inverse square interactions. 
In Appendix \ref{AppendixYoung} some basic facts used in the paper
on the 
irreducible representations of the 
symmetric group are collected.

\section{The Abelian anyon model and non-Abelian Verlinde model}
\label{Verlinde}

\subsection{Abelian case}

Let us  begin with   reviewing the formalism  used in \cite{Notre} 
to calculate the equation of state of Abelian anyons
in the lowest Landau level  of an external magnetic
field. The dynamics of $N$ anyons in the plane
is described by a Aharonov-Bohm  Hamiltonian  in the
background of a constant magnetic field
$B$. 
Here, as in the sequel, the mass $m$ will be  set to 1 and complex coordinates notation  $z_i=x_i + i y_i$,
$\bar z_i
= x_i - i y_i$ will be used.
For our purposes, it will be convenient
to work in the {\em holomorphic} ($+$) or {\em anti-holomorphic} ($-$)
 gauge where the Aharonov-Bohm Hamiltonian 
 takes the form
\beq
H_0^{\pm}= -\sum_i  ( \nabla_i^{\pm} \nablad_i^{\pm} + \nablad_i^{\pm}
\nabla_i^{\pm} )
\label{ham2}
\eeq
The covariant derivatives are defined
as,
\beq
\nabla_i^{\pm} = \derz- i K_{z_i}^{\pm}  \,\,\,\,\nablad_i^{\pm} = \derzd - iK_{\zd_i}^{\pm} 
\eeq
with
\beq
 K_{z_i}^{+} = - i \sum_{j \neq i} \frac{\alpha}{z_i -z_j}  -i \sum_i
b \zd_i  \;\;\;\;  K_{\zd_i}^{+} = 0
\eeq
\beq
 K_{\zd_i}^{-} =  i \sum_{j \neq i} \frac{\alpha}{\zd_i -\zd_j}  + i \sum_i
b z_i  \;\;\;\;  K_{z_i}^{-} = 0
\;\;\;\;,
\eeq
The charge $e$ of each anyon is coupled to the Aharonov-Bohm flux
$\phi$ carried by the other anyons and to the external magnetic field $B$:
therefore the couplings $\alpha=e\phi/2\pi$ and $b=eB/2$.

We are interested in a {\em boson}-based description
of anyons, that is, the Hamiltonian
(\ref{ham2}) is acting  on bosonic wave functions 
$\Psi(z_1,..,z_N; \zd_1,..,\zd_N)$. Then, 
the statistics parameter $\alpha \in [-1,1]$ is such that  
$\alpha=0$ corresponds to bosons and $\alpha= \pm 1$ to fermions.

It is also possible, via a singular
gauge transformation,  to formulate the problem in terms
of a ``free'' Hamiltonian, that is without anyonic Aharonov-Bohm fields,
 at the expense of introducing
multivalued wave functions
\beq\label{free1}
U_{+} H_0^{+} U_{+}^{-1}
 = -\sum_i (\derz \derzd + \derzd \derz  
-2b\zd_i \derzd  -b       )
\eeq
\beq\label{free2}
  U_{-} H_0^{-} U_{-}^{-1}
 = -\sum_i ( \derz \derzd + \derzd \derz  
+2b z_i \derz      +    b         )
\eeq
with $U_{\pm}$ satisfying  the 
equations
\beq
\derz U_{+} =   
 -  \sum_{j \neq i} \frac{\alpha}{z_i -z_j} U_{+}
\;\;\;\;\; \derzd U_{+}=0
\eeq
\beq
\derzd U_{-} =   
   \sum_{j \neq i} \frac{\alpha}{\zd_i -\zd_j} U_{-}
\;\;\;\;\; \derz U_{-}=0
\eeq
which can be considered as the Abelian version of  Knizhnik-Zamolodchikov
equations \cite{KZ},
with solutions
\beq
U_+= \prod_{i<j} (z_i - z_j)^{-\alpha} \;\;\;\;\; 
U_-= \prod_{i<j} (\zd_i - \zd_j)^{\alpha} 
\label{Us}
\eeq
 Note that
due to the multivaluedness of $U$,
$\derz \derzd U \neq \derzd \derz U$.

$H_0^{\pm}$ have singular terms
which arise from 
\beq
\derz ( \frac{1}{\zd_i}) = \derzd (\frac{1}{z_i})= \pi \delta^2(z_i)
\eeq
However, a  potential  accounting for repulsive
contact interactions (which are introduced in order
to implement the exclusion of the diagonal of the
configuration space) and a harmonic well
(which is introduced as a regulator to calculate
thermodynamical properties) can be added to (\ref{ham2})
\beq
V= \sum_{i, j \neq i} \lambda \delta^2(z_i-z_j) +
\sum_{i} \frac{\omega^2}{2} \bar z_iz_i
\;\;\;\;.
\eeq
If we choose
$\lambda= \pi |\alpha|$, the contact term in $V$
cancels exactly the singular terms in $H_0^+$ ($H_0^{-}$) when $\alpha <0$
($\alpha >0$). The total Hamiltonian $H_{\rm AB}=H_0+V$  becomes
\beq
H_{\rm AB}^+ =  -\sum_{i} \left( \derz \derzd + \derzd \derz  
-2b \zd_i \derzd - \frac{\omega^2}{ 2} z_i \zd_i -b \right)
+ 2 \alpha \sum_{i <j } \frac{\derzd -\derzdj }{z_i-z_j}  
\eeq
\beq
H_{\rm AB}^- =  -\sum_{i} \left( \derz \derzd + \derzd \derz  
+2b z_i \derz - \frac{\omega^2}{ 2} z_i \zd_i +b \right)
- 2 \alpha \sum_{i <j } \frac{\derz -\derzj }{\zd_i-\zd_j}  
\eeq
The role of contact interactions  in 
the context of anyons has been extensively discussed in the literature.
Their relevance has been originally stressed
in the study of  soliton solutions in Chern-Simons matter
systems \cite{JaPi} and anyonic wave functions
in the background of an external magnetic field \cite{Gir} (see also
\cite{JoCa}). The issue was subsequently re-considered
by several authors, in  
the perturbative treatment of the Aharonov-Bohm  problem \cite{BeLo}, perturbative
calculations of statistical and thermodynamical quantities \cite{Stat},
self adjoint extensions \cite{Self}, etc. Notice that in the case under
consideration the contact interactions are {\it repulsive}
and that in addition, the orientation of the external
magnetic field and the anyonic flux tubes  are
opposite, a case which do not support solitons \cite{Eza}.
On the other hand, this  is precisely the case considered in \cite{Notre}
since it allows for a physical meaningful lowest Landau level reduction.

Then, if $\omega=0$, the ground state of $H_{\rm AB}^+$, which corresponds
to $b>0$ and $\alpha<0$, is given by analytic wave functions while
the ground state of $H_{\rm AB}^-$, associated to  $b<0$ and
$\alpha >0$, is given by anti-analytic wave functions. 

As we mentioned before, the harmonic attraction lifts the
degeneracy of the ground state. In order to see this, it is
convenient to re-define the wave functions as
\beq
\Psi^{\pm}(z_1,..,z_N;\zd_1,..,\zd_N)=\prod_i \exp(-\frac{\omega_t\mp b}{2} z_i \zd_i) 
\psi^{\pm}(z_1,..,z_N;\zd_1,..,\zd_N)
\label{rede}
\eeq
where $\omega_t\equiv
\sqrt{\omega^2+ b^2}$.  
The Hamiltonian acting on $\psi^{\pm}$ is then
\bea
{H}^{+} &=&  -\sum_{i} \left( \derz \derzd + \derzd \derz -
(\omega_t+ b) \zd_i  \derzd   
-(\omega_t- b) z_i \derz  
-\omega_t \right)
\nonumber \\
& &
 + 2\alpha  \sum_{i<j} (  \frac{\derzd- \derzdj}{z_i-z_j} -\frac{\omega_t- b}{2} )  
\label{abe1}
\eea
\bea
{H}^{-} &=&  -\sum_{i} \left( \derz \derzd + \derzd \derz 
-(\omega_t+ b) \zd_i \derzd  
-(\omega_t- b) z_i  \derz   
-\omega_t \right)
\nonumber
\\
& & - 2\alpha  \sum_{i<j} ( \frac{\derz- \derzj}{\zd_i-\zd_j}  -\frac{\omega_t+ b}{2})  
\label{abe2}
\eea

Acting on analytic and anti-analytic wave functions
\beq
\psi^+= \prod_i z_i^{\ell_i} \;\;\;\;\;\;\; \psi^-= \prod_i \zd_i^{\ell_i}
\eeq
the Hamiltonians ${H}^{\pm}$  have the spectrum
\beq
E_N= N\omega_t +  (\sum_i \ell_i - \frac{\alpha}{2}N(N-1))(\omega_t- b)
\eeq
\beq
E_N= N\omega_t +  ( \sum_i \ell_i + \frac{\alpha}{2} N (N-1))(\omega_t+ b)
\eeq
Notice that the exclusion of the diagonal of the configuration space, 
in view of (\ref{Us}) and our discussion
on the sign of $\alpha$, is realized for the wave functions in the
singular (s) gauge
\bea
\psi^{s +} &=&  \prod_{i<j} (z_i - z_j)^{-\alpha}  \prod_i z_i^{\ell_i} 
\exp(-\frac{\omega_t- b}{2} z_i \zd_i) \\    
\psi^{s -} &=& \prod_{i<j} (\zd_i - \zd_j)^{\alpha}  
\prod_i \zd_i^{\ell_i} 
\exp(-\frac{\omega_t+ b}{2} z_i \zd_i)    
\eea
which do vanish at coinciding points. 

\subsection{Non-Abelian case: the Verlinde model}

Let us generalize the above construction to
SU(2) non-Abelian Chern-Simons  particles
in the lowest Landau level of an external Abelian
magnetic field. We are considering the case where
 $N$ identical {\it bosonic} particles  are in the isospin $1/2$
{\it fundamental} representation of $SU(2)$. The wave
functions of the system $\Psi(z_1,..,z_N,\bar z_1,\ldots ,\bar z_N)$ 
belongs to the tensor product $\Gamma_1 \times ...\times \Gamma_N$
where $\Gamma_i$ is the two-dimensional space in which the
SU(2) generators,
\beq
T^a_i= \frac{\sigma^a_i}{2} \,\,\,\,\, a=1,2,3 \,\,\,\,\,\,
 i=1,..,N
\eeq
act. The non-Abelian generalization of the Aharonov-Bohm
 Hamiltonian in the holomorphic and anti-holomorphic gauges is given by,
\beq
H_0^{\pm}= -\sum_i  ( \nabla_i^{\pm} \nablad_i^{\pm} + \nablad_i^{\pm}
\nabla_i^{\pm} )
\label{ham3}
\eeq
where we the covariant derivatives are defined as before but with non
Abelian gauge fields,
\beq
 K_{z_i}^{+} =  ig \sum_{j \neq i} \frac{T_i^aT_j^a}{z_i -z_j}  -i \sum_i
b \zd_i  \;\;\;\;  K_{\zd_i}^{+} = 0
\eeq
\beq
 K_{\zd_i}^{-} =  -ig \sum_{j \neq i} \frac{T_i^a T_j^a}{\zd_i -\zd_j}  
+ i \sum_i
b z_i  \;\;\;\;  K_{z_i}^{-} = 0
\;\;\;\;,
\eeq
Here $g$ is the Chern-Simons coupling constant ($g=1/2\pi \kappa$
in \cite{LeePRL}) and satisfies the topological constraint $g=2/n$ 
with $n$ an integer.

Naturally, other choices such as Coulomb or axial
gauges are possible. The connection among these
different gauge choices  is obtained via 
$W^{-1} H_0 W$ where $W$ is not necessarily unitary. Unlike
the Abelian case where $W$ is known, for the non-Abelian
case an explicit expression is available only in the $2$-body
case. 

The Hamiltonian acts on totally symmetric wave
functions with bosonic interchange conditions, meaning
that the wave function is symmetric under the interchange
of {\it both} coordinates and isospin indices,
\beq
\Psi_{i_1,..,i_m,..,i_l,..,i_N}(z_1,..,z_m,..,z_l,..,z_N) =
\Psi_{i_1,..,i_l,..,i_m,..,i_N}(z_1,..,z_l,..,z_m,..,z_N) 
\eeq
As in the Abelian case, there exists a Hamiltonian without anyonic
gauge fields but acting on
wave functions with {\it non trivial} boundary conditions.
This Hamiltonian, $U_{\pm} H_0^{\pm} U^{-1}_{\pm}$, takes  the same
form than in the Abelian case where  $U_{\pm}$ satisfies  the
non-Abelian Knizhnik-Zamolodchikov  equations

\beq
\derz U_{+} =   U_{+}   g \sum_{j \neq i} \frac{T_i^aT_j^a}{z_i -z_j}  
\quad \;\;\; \derzd U_{+}=0
\eeq
\beq
\derzd U_{-} = -  U_{-}  g \sum_{j \neq i} \frac{T_i^a T_j^a}{\zd_i -\zd_j}
\quad \;\;\; \derz U_{-}=0 
\eeq

As in the Abelian case, the Hamiltonians $H_0^{\pm}$ have singular terms
that can be eliminated by adding an appropriate  potential
$V$. Thus, as a non-Abelian generalization,
 we are led to consider
$
H_{\rm AB}^{\pm}= H_0^{\pm} + V
$
with 
\beq
V=
  |g| \pi \sum_{i,j \neq i} T_i^a T_j^a \delta^2(z_i-z_j) +
\sum_{i} \frac{\omega^2}{2} \bar z_i z_i
\eeq
It can be shown that this potential corresponds
to a repulsive interaction in the bosonic sector. We then  obtain
\beq
H_{\rm AB}^+ =  -\sum_{i} \left( \derz \derzd + \derzd \derz  
-2b\zd_i \derzd - \frac{\omega^2 }{ 2} z_i \zd_i -b \right)
- 2 g \sum_{i<j} T_i^a T_j^a\frac{\derzd -\derzdj   }{z_i-z_j}  
\eeq
\beq
H_{\rm AB}^- =  -\sum_{i} \left( \derz \derzd + \derzd \derz  
+2b z_i \derz - \frac{\omega^2 }{ 2} z_i \zd_i +b \right)
+ 2 g \sum_{i<j} T_i^a T_j^a\frac{\derz-\derzj}{\zd_i-\zd_j} 
\eeq
and factorizing the Gaussian factor in the wavefunctions as in (\ref{rede})
\bea 
{H}^{+} &=&  -\sum_{i} \left( \derz \derzd + \derzd \derz -
(\omega_t+ b) \zd_i  \derzd   
-(\omega_t- b) z_i \derz  
-\omega_t \right) 
\nonumber\\
& &
 - 2g  \sum_{i<j} T_i^aT_j^a(  \frac{\derzd- \derzdj}{z_i-z_j} 
-\frac{\omega_t- b}{2} )  
\label{nabe1}
\eea
\bea
{H}^{-}  &=&  -\sum_{i} \left( \derz \derzd + \derzd \derz 
-(\omega_t+ b) \zd_i \derzd  
-(\omega_t- b) z_i  \derz   
-\omega_t \right)
\nonumber \\ 
& &+2g  \sum_{i<j}T_i^aT_j^a  \frac{\derz- \derzj}{\zd_i-\zd_j}  
-\frac{\omega_t+ b}{2}) 
\label{nabe2} 
\eea

Acting on analytic (anti-analytic) wave functions, 
${H}^{+}$ (${H}^{-}$) takes  the form,
\beq
{H}^+= N \omega_t + 
\left(\sum_{i} \ell_i + \hat\Omega_{I,N}\right) (\omega_t- b)
\eeq
\beq
{H}^-= N \omega_t + 
\left( \sum_{i} \ell_i - \hat\Omega_{I,N}\right) (\omega_t+ b)
\eeq
where 
\beq
\hat{\Omega}_{I,N} =  g\sum_{ i<j} T_i^a T_j^a
\eeq 
It can be easily shown that the operator $\hat{\Omega}$ has eigenvalues,
\beq
\Omega_{N,I} = \frac{g}{2}\left( I(I+1) - \frac{3}{4} N\right)
\label{Omega1}
\eeq
where $N$ is as before the total number of particles
and $I$ is the total isospin.

For analytic (anti-analytic) wave functions,  
the  spectrum reads
\be
E_{N,I} = N \omega_t + 
         \left(
	 \sum_{j=1}^{N} \ell_j  +\Omega_{N,I}\right)(\omega_t - b) 
\label{E-NI}\ee
\be
E_{N,I} = N \omega_t + 
          \left(
	 \sum_{j=1}^{N} \ell_j -\Omega_{N,I} \right)(\omega_t + b)
\label{}\ee

\section[]{The Abelian and non-Abelian anyon models revisited}
\label{revisited}

In the standard presentation given above of the Abelian  and non
Abelian Verlinde anyon models, short range $\delta^2(z_i-z_j)$ 
interactions have been
added by hand to the
 non hermitian Hamiltonians expressed in the holomorphic or anti-holomorphic 
 gauges.
We now present another  approach  to derive the same
Hamiltonians (\ref{nabe1},\ref{nabe2}), which starts
from free Pauli Hamiltonians, therefore 
implying spin coupling 
 to the local magnetic field carried by the anyons. In 
 addition, 
the eigenstates redefinitions with  respect to the
vortex (short distance) and the magnetic field (long distance) will be
treated on an equal footing.

In the singular gauge, let us start with the free $N$-body Pauli 
Hamiltonians
\be\label{1} H_{\rm free}^{+}=-2\sum_{i=1}^N\partial_i\bar\partial_i\ee
\be\label{2} H_{\rm free}^{-}=-2\sum_{i=1}^N\bar \partial_i\partial_i\ee
where the index $\pm$ refers here to the spin degree of freedom.

If one  considers the additional coupling to an external magnetic field, 
then, in the symmetric gauge,  
$\partial\to\partial-b\bar z/2$ and
$\bar\partial\to\bar\partial+bz/2$.
Of course one could choose other gauges for the $B$ field, 
as the holomorphic or
anti-holomorphic gauges, in which case in (\ref{1}) $\partial\to\partial-b\bar
z,\bar \partial\to\bar \partial$, and in (\ref{2}), $\bar\partial\to
\bar\partial+bz, \partial\to \partial$, which yield basically the
Hamiltonians (\ref{free1},\ref{free2}) discussed above.  
However, we insist at this point
on using the symmetric gauge, since it is the natural gauge to work 
with in the presence of the singular Aharonov-Bohm flux tubes.

Indeed, the anyon model is defined via the non 
trivial monodromy of the $N$-body
eigenstates of $H_{\rm free}$ 
\be\label{3} \psi_{\rm free}(z_1,z_2,\ldots ,z_N;
\bar z_1,\bar z_2,\ldots ,\bar z_N)=e^{-i\alpha\sum_{k<
l}\theta_{kl}}\Psi(z_1,z_2,\ldots ,z_N; \bar z_1,\bar z_2,\ldots ,\bar z_N)\ee
where $\sum_{k< l}\theta_{kl}$ is the
sum of the relative angles between  pairs of particles. As said before,
 $\psi(z_1,z_2,\ldots ,z_N; \bar z_1,\bar z_2,\ldots ,\bar z_N)$ is by
 convention bosonic in the regular gauge
 with the statistics parameter $\alpha=0$ 
for Bose statistics, 
and $\alpha=\pm 1$ for Fermi statistics.

Looking at (\ref{3}) as a singular gauge transformation,
one would obtain, in the symmetric gauge,
 a $N$-anyon  Aharonov-Bohm Hamiltonian in the background 
 of the external magnetic field 
with  $\mp\pi\alpha\sum_{i<j} \delta^2(z_i-z_j)$
interactions and $ \mp \sum_i b$  shifts, induced by  the
spin up or spin down coupling to the local magnetic field of the
vortices and the homogeneous background magnetic field. The parameter
$\alpha$ represents as usual the Aharonov-Bohm flux carried by the anyons 
in units of the quantum of flux.

The short range (contact) interactions  have to implement
the exclusion of the diagonal of the
configuration space (Pauli exclusion), and thus have to
be repulsive. So, depending
of the sign of $\alpha$, the spin up Hamiltonian (\ref{1}) $(\alpha\in
[-1,0])$
or spin down Hamiltonian
(\ref{2}) $(\alpha\in[0,1])$ have
to be used.

To materialize  the short range repulsion in the eigenstates,
one  proceeds by
 redefining
\be\label{4} \Psi(z_1,z_2,\ldots ,z_N; \bar z_1,\bar z_2,\ldots ,\bar z_N)=
\prod_{i<j}|z_i-z_j|^{\mp \alpha}\tilde\psi(z_1,z_2,\ldots ,z_N; \bar
z_1,\bar z_2,\ldots ,\bar z_N)\ee
This is nothing but saying that the eigenstates do vanish as
quickly as $r_{ij}^{\mp \alpha}$ when  particles $i$ and $j$ come close 
together (again the $\mp$ sign has been chosen
accordingly to the sign of $\alpha$).
At this point,   (\ref{3},\ref{4})  together give
\be\label{5}\psi_{\rm free}=
\prod_{k<l}( z_k- z_l)^{-\alpha}\tilde\psi_{+}\quad\quad\alpha\in [-1,0]
\ee
\be\label{6}
\psi_{\rm free}=\prod_{k<l}(\bar z_k-\bar z_l)^{\alpha}\tilde\psi_{-}\quad\quad \alpha\in [0,1]\ee
where both $\prod_{k<l}( z_k- z_l)^{-\alpha}$ and $  \prod_{k<l}
(\bar z_k-\bar z_l)^{\alpha}$ 
are solutions of the holomorphic and anti-holomorphic 
Knizhnik-Zamolodchikov equations in the Abelian case. 
The  non hermitian Hamiltonian acting on $\tilde{\psi}$  rewrites 
$$\tilde {H}^{+}=   -2  \sum _{i=1}^{N} 
              \left[ \partial_i\bar\partial_i 
              -({b\over 2})^2 z_i\bar z_i
              +{b\over 2}(z_i\partial_i-\bar z_i\bar\partial_i)\right]
              \quad\quad\quad\quad\quad\quad $$
\be\label{7}  +2\alpha\sum_{i<j}
({\bar\partial_i-\bar\partial_j\over z_i-z_j}+{b\over 2})
- \sum_i{b}
\ee

$$\tilde {H}^{-}=   -2  \sum _{i=1}^{N} 
              \left[ \bar\partial_i\partial_i 
              -({b\over 2})^2 z_i\bar z_i
              +{b\over 2}(z_i\partial_i-\bar z_i\bar\partial_i)\right]
              \quad\quad\quad\quad\quad\quad $$
\be\label{8}  -2\alpha\sum_{i<j} ({\partial_i-\partial_j\over \bar z_i-
\bar z_j}-{b\over 2})
             + \sum_i{b}
\ee

The external magnetic field did not yet plaid any role in the eigenstates
redefinition, thus the Hamiltonians (\ref{7}, \ref{8}) expressed in the
holomorphic or anti-holomorphic gauges with respect to the vortices, but 
in the symmetric gauge with respect to the external magnetic
field. Also, there is no such singular gauge as (\ref{3}) for a
homogeneous
magnetic field.
 Let us however extract from the eigenstates the Landau exponential
factor
$\exp(\pm {b\over 2}\sum_iz_i\bar z_i)$. It should be considered
on the same footing as $\prod_i|z_i-z_j|^{\pm\alpha}$ in (\ref{4}),  as one
can easily realize by considering 
the 2-dimensional identity
\be\int dr'\ln|\vec r-\vec r'|=\pi r^2/2\ee
It means that  a magnetic field  can be regarded as the
average of a distribution of vortices \cite{Desbois}.  Let us redefine
\be\label{9}\psi_{\rm free}=\prod_{k<l}( z_k- z_l)^{-\alpha}\exp(-{1\over
2}b\sum_iz_i\bar z_i){\psi}_{+}\ee
\be\label{10}\psi_{\rm free}=\prod_{k<l}(\bar z_k-\bar z_l)^{\alpha}\exp({1\over
2}b\sum_iz_i\bar z_i){\psi}_{-}\ee
where the prefactors  $\psi_{\rm free}=U^{\pm}\psi_{\pm}$ are now solutions of the
holomorphic and anti-holomorphic 
Knizhnik-Zamolodchikov  equations in presence of the external $B$ field
\be\label{11} ( \partial_i +{b\over
2}\bar z_i+\alpha\sum_{j\ne i}{1\over z_i-z_j})U^+=0 \quad  (\bar\partial_i +{b\over
2}z_i)U^+= 0\ee

\be\label{12} (\bar \partial_i -{b\over
2} z_i-\alpha\sum_{j\ne i}{1\over \bar z_i-\bar z_j})U^-=0 \quad ( \partial_i -{b\over
2}\bar z_i)U^-=0\ee

In order for (\ref{9},\ref{10}) to be physically meaningful, i.e. short
distance 
vanishing eigenstates due to the repulsive vortices, and  
long distance exponential damping due to the magnetic field, 
one concludes that in (\ref{9}), $\alpha<0, b>0$, 
whereas in (\ref{10}),  $\alpha>0, b<0$. In both cases,
the local vortices carried by the anyons are antiparallel to the external
magnetic field, i.e. a screening regime.
The resulting  Hamiltonian in the holomorphic and anti-holomorphic gauges
acting on $\psi_{\pm}$ narrows down to
\be\label{13}{H^+} = -2  \sum _{i=1}^{N} 
              \left[ \partial_i\bar\partial_i 
                -b\bar z_i\bar \partial_i\right]
+2\alpha\sum_{i<j}{\bar \partial_i-\bar \partial_j\over  z_i- z_j}\ee
\be\label{14}{H^-} = -2  \sum _{i=1}^{N} 
              \left[ \bar\partial_i\partial_i 
                +{b}z_i\partial_i\right]
-2\alpha\sum_{i<j}{\partial_i-\partial_j\over \bar z_i-\bar z_j}\ee

 Any analytic function of the variables $z_i$ is an eigenstate of
 the holomorphic Hamiltonian (\ref{13}), whereas a anti-analytic function of
 $\bar z_i$ is an eigenstate of (\ref{14}). From (\ref{9},\ref{10}), one infers 
  the infinitely degenerate  ground state with zero energy\footnote{The lowest Landau level
  spectrum has been shifted downward by the spin induced shift.}
  \be\label{15}
 \psi_{\rm free}=\prod_{i<j} (z_i-z_j)^{-\alpha}\prod_iz_i^{\ell_i}
 \exp(-{1\over 2}b\sum_iz_i \bar z_i),  \quad\quad \ell_i \ge 0
\ee
\be\label{16} \psi_{\rm free}=\prod_{i<j} (\bar z_i-\bar z_j)^{\alpha}\prod_i\bar z_i^{\ell_i}
 \exp({1\over 2}b\sum_iz_i \bar z_i),  \quad\quad \ell_i \ge 0
\ee
where the orbital quantum numbers $\ell_i=0,1,...\infty$ are such that the
eigenstates are symmetric. If one leaves aside the
anyonic prefactors $\prod_{i<j} (z_i-z_j)^{-\alpha}$ and
$\prod_{i<j} (\bar z_i-\bar z_j)^{\alpha}$, the $N$-anyon
ground state is a symmetrised product of $1$-body Landau ground states with
orbital angular momentum $\ell_i$, as for ideal one-dimensional bosons.

 To compute its thermodynamical properties, one has to
 regularize the system at long distance.
 Adding  a harmonic well confinement $\sum_i{1\over 2}\omega^2\bar z_i z_i$
 to the Hamiltonians (\ref{1},\ref{2}),  an $N$-body anyonic
eigenstate in the lowest Landau level
of an external magnetic field  is
still entirely characterized by  a product
 $1$-body eigenstates with a given orbital quantum number
$\ell_i$.

If $b>0$, $b=+\omega_c$
where $\omega_c$ is is half the cyclotron frequency, and
  $\alpha\in [-1,0]$. If $b<0$, $b=-\omega_c$, and $\alpha\in [0,1]$.
In the presence of the  harmonic well, 
the eigenstates  are  still given by (\ref{15},\ref{16}), but now
with $\omega_{\rm c}\to\omega_{\rm t}=\sqrt{\omega_{\rm c}^2+\omega^2}$.
It follows that (\ref{9},\ref{10}) should rewrite as
\be
\label{17}
\psi_{\rm free}=\prod_{k<l}( z_k- z_l)^{-\alpha}\exp(-{1\over
2}\omega_t\sum_iz_i\bar z_i)\psi_{+}\ee
\be\label{18}\psi_{\rm free}=\prod_{k<l}( \bar z_k- \bar z_l)^{\alpha}\exp(-{1\over
2}\omega_t\sum_iz_i\bar z_i)\psi_{-}\ee
to get the holomorphic and anti-holomorphic Hamiltonians in the presence of
the harmonic well
\bea
{H^+} &=& -2  \sum _{i=1}^{N} 
              \left[ \partial_i\bar\partial_i 
                -{\omega_t+\omega_c\over 2}\bar z_i\bar \partial_i
		-{\omega_t-\omega_c\over 2} z_i \partial_i
		\right]
\nonumber \\
& & +2\alpha\sum_{i<j}\left[{\bar \partial_i-\bar \partial_j\over  z_i-
z_j}-{\omega_t-\omega_c\over 2}\right]+\sum_i(\omega_t-\omega_c)
\label{19}
\eea
\bea
{H^-} &=& -2  \sum _{i=1}^{N} 
              \left[ \partial_i\bar\partial_i 
                -{\omega_t-\omega_c\over 2}\bar z_i\bar \partial_i
		-{\omega_t+\omega_c\over 2} z_i \partial_i
		\right]
\nonumber \\
& &-2\alpha\sum_{i<j}\left[{ \partial_i- \partial_j\over \bar z_i-
\bar z_j}-{\omega_t-\omega_c\over 2}\right]+\sum_i(\omega_t-\omega_c)
\label{20}
\eea
These Hamiltonians are identical, up to a constant energy shift, to the
Hamiltonians (\ref{abe1},\ref{abe2}) obtained above since,
when acting 
on regular wavefunctions,  $\partial_i\bar\partial_i=\bar\partial_i\partial_i$.

The $N$-body spectrum is nothing else but the sum of the $1$-body spectra
$\epsilon_{\ell_i}=(\omega_t-\omega_c)+\ell_i
(\omega_{\rm t}
-\omega_{\rm c})$, shifted by the $2$-body statistical energy $-{1\over2}N(N-1)(\omega_{\rm t}
-\omega_{\rm c})\alpha$.
So, the virtue of the harmonic confinement has been to partially lift the
degeneracy with respect to the $\ell_i$'s, and
to dress the spectrum with an explicit  $\alpha$ dependence
\be\label{21} E_N=N(\omega_t-\omega_c)
+(\sum_i
\ell_i
\mp{1\over2}N(N-1)\alpha)(\omega_t-\omega_c)\ee

Let us now turn to the non-Abelian case :  
The non-Abelian Knizhnik-Zamolodchikov  equations become in place of (\ref{11}) and (\ref{12})
\be\label{22} ( \partial_i +{b\over
2}\bar z_i)U^+-U^+g\sum_{j\ne i}{T_i^aT_j^a\over z_i-z_j}
=0 \quad (\bar \partial_i +{b\over
2}z_i)U^+= 0\ee

\be\label{23} (\bar \partial_i -{b\over
2}z_i)U^-+U^-g\sum_{j\ne i}{T_i^aT_j^a\over \bar z_i-\bar z_j}
=0 \quad ( \partial_i -{b\over
2}\bar z_i)U^-=0\ee
One can proceed exactly in the same way as in the Abelian case, i.e. 
 start from the free Pauli Hamiltonians $H^{\pm}_{\rm free}$ in
(\ref{1},\ref{2}), eventually coupled to the external magnetic field,
and redefine the free eigenstates according to
\be\label{24} \psi_{\rm free}(z_1,z_2,\ldots ,z_N;
\bar z_1,\bar z_2,\ldots ,\bar z_N)=
U^{\pm}\psi_{\pm}(z_1,z_2,\ldots ,z_N; \bar z_1,\bar z_2,\ldots ,\bar z_N)\ee
The  non Hermitian Hamiltonian,  in the presence of an
harmonic well rewrites in the holomorphic and anti-holomorphic gauges as
\bea
{H^+} &=& -2  \sum _{i=1}^{N}
              \left[ \partial_i\bar\partial_i
                -{\omega_t+\omega_c\over 2}\bar z_i\bar \partial_i
		-{\omega_t-\omega_c\over 2} z_i \partial_i
		\right]
\nonumber \\
& &
-2g\sum_{i<j}T_i^aT_j^a\left[{\bar \partial_i-\bar \partial_j\over  z_i-
z_j}-{\omega_t-\omega_c\over 2}\right]+\sum_i(\omega_t-\omega_c)
\label{25}
\eea
\bea
{H^-} &=& -2  \sum _{i=1}^{N}
              \left[ \partial_i\bar\partial_i
                -{\omega_t-\omega_c\over 2}\bar z_i\bar \partial_i
		-{\omega_t+\omega_c\over 2} z_i \partial_i
		\right]
\nonumber \\
& &
+2g\sum_{i<j}T_i^aT_j^a\left[{ \partial_i- \partial_j\over 
\bar z_i-
\bar z_j}-{\omega_t-\omega_c\over 2}\right]+\sum_i(\omega_t-\omega_c)
\label{26}
\eea
a generalization of (\ref{19},\ref{20}) which coincides with
(\ref{nabe1},\ref{nabe2}).
Accordingly the spectrum of (\ref{25}) (respectively(\ref{26})) acting on analytic
(respectively anti-analytic) eigenstates is
\be\label{27} E_{N,I}=N(\omega_t-\omega_c)
+(\sum_i
\ell_i
\pm \Omega_{N,I})(\omega_t-\omega_c)\ee
where $\Omega$ has been defined in (\ref{Omega1}). 

The cases $b>0,\alpha\in[-1,0]$ and
$b<0,\alpha\in[0,1]$ are equivalent. So in the sequel one will assume 
without loss of generality $b\equiv \omega_c>0, \alpha\in[-1,0]$, i.e. the holomorphic gauge.

\section{The thermodynamics of two non-Abelian Chern-Simons particles 
in an external magnetic field}

\label{2-body}

\subsection{Virial expansion}

In this section, we study the thermodynamic quantities in which anyonic
statistics manifests itself. For this purpose 
the virial coefficients $a_k$, which result from   
the expansion of the pressure $P$ in terms
of the particle density $\rho$, can be used
\be
{1\over V}\ln \Xi= \beta P = \sum_{k=1}^{\infty} a_k {\rho}^k 
\label{virial}\ee
Using the cluster expansion,
\be
\ln \Xi = \sum_{k=1}^{\infty} b_k z^k \;,
\label{cluster}\ee
where $z = e^{\beta \mu}$ is the fugacity, it is possible to express 
the $k$-th virial coefficient
in terms of partition functions $Z_j$ with $j\leq k$. 
Indeed, from
\be \Xi = \sum_{N=0}^{\infty} Z_N z^N \ee
it follows that,
\bea
b_1 &=& Z_1\;, \nonumber \\  
b_2 &=& Z_2 -\frac12 Z_1^2\;, \nonumber \\  
b_3 &=& Z_3 -Z_2 Z_1 + \frac13 Z_1^3 \;,
\label{b}\eea
and
\bea
\tilde a_1 &=& 1 \;, \nonumber \\
\tilde a_2 &=& - \tilde b_2 \;, \nonumber \\
\tilde a_3 &=& -2 \tilde b_3 + 4 \tilde b_2^2 \;,
\label{a}\eea 
where $\tilde a_k =a_k/V^{k-1} $ and  $\tilde b_k =b_k /b_1^k$. 

Considering a harmonic potential as a regulator, the thermodynamic
quantities
in a box of infinite volume $V$  can be calculated from those in a harmonic well
of frequency $\omega$
by using the following ``thermodynamic limit prescription'': in the
thermodynamic limit,
the cluster coefficients  $b_k$ are obtained from those in the
harmonic well $b_k^{\omega}$ by
(depending on the cluster coefficient order $k$) \cite{regularization}
\be
\frac{1}{(k \beta^2 \omega^2)^{d/2}} \to \frac{V}{\lambda_T^d} \;,
\label{TDprescr}
\ee
where $d$ is the spatial dimension.  
In two dimensions, in the presence of an external magnetic field, 
(\ref{TDprescr}) becomes
\be
\frac{1}{k \beta ( \omega_t - \omega_c) } \to {V}{\rho_L} \;,
\label{TDprescrB}
\ee
where $\rho_{L}={eB\over 2\pi}$ is the Landau level degeneracy per unit volume.

\subsection{2-body Abelian anyon case}

Let us consider two non-Abelian Chern-Simons particles in an external 
magnetic field. This problem can be solved exactly since it decomposes 
into two 2-body Abelian anyon problems. We start by discussing the Abelian 
anyon problem paying special attention to the peculiarities of the 
  strong magnetic field limit which are important for understanding 
the non-Abelian case. 
   
The spectrum for the relative motion of
two Abelian anyons in
an external magnetic field $B$
and a harmonic well $\omega$ is
\be\label{30} E_{nm}=(2n+1+|m-\alpha|)\omega_t-(m-\alpha)\omega_c \;, \ee 
\be\label{31} \psi_{nm}=e^{i(m-\alpha)\theta}r^{|m-\alpha|}L_n(\omega_t z\bar z)\;.\ee
The spectrum and eigenstates are  periodic with
period 2 in $\alpha$ since $m$  has to be chosen to
 be an even (odd) integer for boson (fermion) based anyons. So 
one can always restrict $\alpha\in[-1,1]$.

In the thermodynamic limit $\omega\to 0$, 
the spectrum and eigenstates  rewrite as
\be\label{32} m\ge \alpha : \quad E_{nm}=(2n+1)\omega_c\ee
\be\label{34} \psi_{nm}=z^{m-\alpha}L_n(\omega_c z\bar z)\ee
or
\be\label{33} m\le \alpha : \quad E_{nm}=(2n+1+2(\alpha-m))\omega_c\ee
\be\label{35} \psi_{nm}=\bar z^{\alpha-m}L_n(\omega_c z\bar z)\ee

The projection onto the lowest Landau level corresponds to
$n=0$. This implies that the wave functions
are  
analytic (anti-analytic) for $m \ge \alpha$ ($m \le \alpha$).
 Notice nevertheless that the projection is not well defined at the bosonic 
end (when $\alpha \to 0^+$ for boson based anyons or 
 $\alpha \to -1^+$ for fermion based anyons). Indeed, let 
us look at the ground state $n=0$ in the the boson based description. 
If $\alpha \in [-1,0]$, the analytic ground state basis (\ref{34}) is 
complete since then the $m=0$ state belongs to this basis.   
However, if  $\alpha\in [0,1]$, the analytic 
ground state basis is incomplete since the $m=0$ state is anti-analytic and
has the  energy which varies linearly with $\alpha$, 
joining the ground state basis when $\alpha  \to 0^+$, 
 as represented in Fig. 1.

\begin{figure}
\centerline{\hbox{
   \epsfig{figure=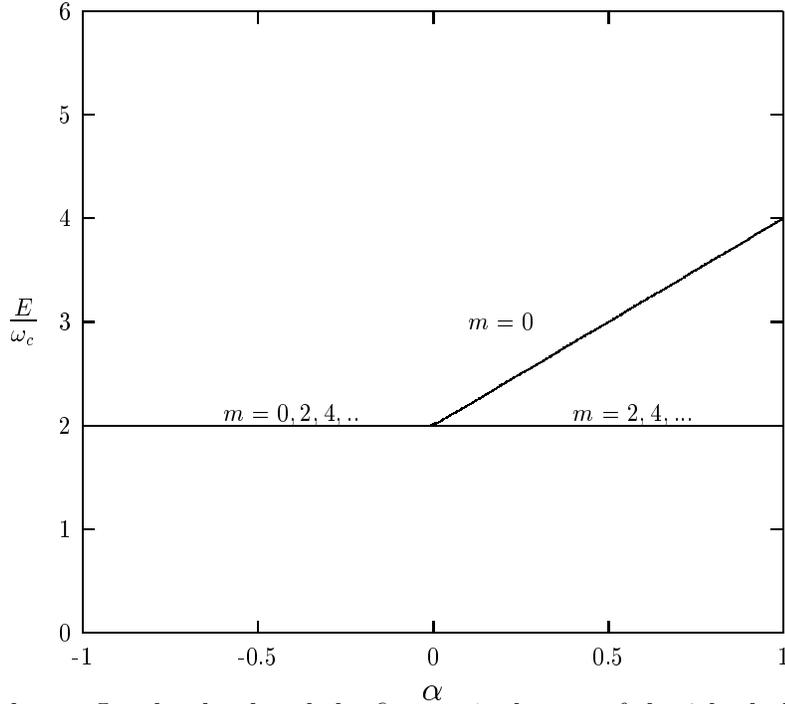,width=11cm}}
 }
 \caption{The lowest Landau level and the first excited state of the 2-body boson 
 based anyon problem.}
 \label{phasediagram}
\end{figure}

The 2-anyon second virial coefficient in a magnetic field 
has been computed in
\cite{JoCa} 
($b$ and $f$ stand for boson based and fermion based anyons, respectively)
\be\label{36} a_2^{b,f}= { {\scriptstyle \lambda_T^2 }\over{\scriptstyle
               x}}\bigg( \mp{{\scriptstyle 1}\over {\scriptstyle 4}} \tanh x \,
               -{{\scriptstyle 1}\over {\scriptstyle 2}}\alpha \,
              -{{\displaystyle e^x (e^{-2x\alpha} -1)}\over
             {\scriptstyle 4}}{( {{\scriptstyle 1}\over
               {\scriptstyle \sinh x}}
             \pm{{\scriptstyle 1}\over{\scriptstyle \cosh x}}) }
             +{{\scriptstyle 1\pm 1}\over {\scriptstyle 2}}
           (e^{x(|\alpha| -\alpha)}-1) \bigg) \quad\quad\quad\quad
	   \ee
where $x=\beta\omega_c$
and $\lambda_T=\sqrt{2\pi\beta}$ is the thermal wavelength. It is depicted in the Fig. 2 
both for low magnetic fields (low $x$) and for high magnetic fields 
(large $x$).

In the  limit of the strong magnetic field, one obtains 
for boson based anyons  
\be
a_2^{b} = {1\over 2\rho_L}( -1 - 2 \alpha ) \;\;\; \alpha\in[-1,0] \; , 
\ee
\be
a_2^b = {1\over 2\rho_L} \left( - 1 - 2 \alpha 
+ 4(1-e^{-2\alpha x}) \right) \;\;\; \alpha\in[0,1] ,  
\ee

\begin{figure}
\centerline{\hbox{
   \epsfig{figure=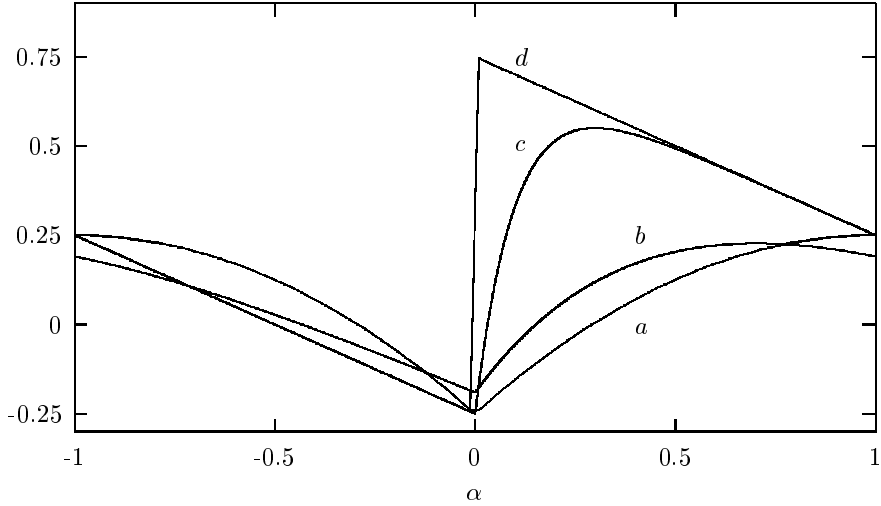,width=13cm}}
 }
 \caption{Second virial coefficient as a function of the
statistics parameter $\alpha$:   (a) $\frac{a_2}{\lambda_T^2}$ for zero 
magnetic field, 
(b) $\frac{a_2 \rho_L}{2}$ for $x=1$,   
(c) $ \frac{a_2 \rho_L}{2}$ for $x=5$,  
(d) $\frac{a_2 \rho_L}{2}$ for $x\to \infty$.}
 \label{v2ab}
\end{figure}

For fermion based anyons, the resulting expression is 
\be\label{toto}
a_2^f  = {1\over 2\rho_L}\left( 1 - 2 \alpha  -4 e^{-2x(1+\alpha)} \right) \;, 
\label{alpha-}\ee
As the exponential factor is only relevant for $\alpha \to -1$, (\ref{toto}) 
can be rewritten as,
\be
a_2^f  = {1\over 2\rho_L}\left( 1 - 2 \alpha  -4 e^{-2x(1+\alpha)} \right) \;, 
\label{alpha--} \;\;\; \alpha\in[-1,0] \; , \ee
\be
a_2^f = {1\over 2\rho_L}( 1 -2\alpha ) \;,
\label{alpha+}\;\;\; \alpha\in[0,1] ,   \ee

Note that as far as $x$ is finite, the virial coefficients are
continuous functions of $\alpha$. These results should
be compared to the ones that would arise if one projects
into the lowest Landau level at the very beginning,
\be
a_2^{b} = {1\over 2\rho_L}( -1 - 2 \alpha ) \;\;\; \alpha\in[-1,0] \; , 
\ee
\be
a_2^b = {1\over 2\rho_L} \left( - 1 - 2 \alpha 
+ 4 \right) \;\;\; \alpha\in[0,1] ,  
\ee
\be
a_2^f = {1\over 2\rho_L}( 1 -2\alpha ) \;,
\label{alphaLLL}\;\;\;    \ee
$a_2^{b}$ has now a jump at $\alpha=0$ while
$a_2^f(-1) \neq a_2^f(1)$. These discontinuities are  a  direct consequence of
the projection onto the lowest Landau level
which ignores  the $m=0$ ($m=1$) state leaving the ground state
-or in other words the vanishing  of the gap- in the
boson (fermion) based description for $\alpha=0$
($\alpha=-1$). These states, when properly taken into account, smooth the
discontinuities as can be seen in Fig. 2.

\subsection{2-body non-Abelian problem}

Consider now the $2$-body non-Abelian SU(2) case.
The isospin is either  $I=1$ (triplet)
\beq
\chi^{}_{1,1} =  |+ +>  \;\;\;\;\; 
\chi^{}_{1,0} = \frac{1}{\sqrt{2}}\, ( |+ -> + |- +> )   \;\;\;\;\;
\chi^{}_{1,-1} = |- -> 
\eeq
or $I=0$ (singlet) 
\beq
\chi^{}_{0,0} = \frac{1}{\sqrt{2}}\, ( |+ -> - | - +> ) \; . 
\eeq

The  wave function  is
\beq
\psi^{}_{1,m_I} =\phi^{s}_{m_I} (z_1,z_2)\, \chi^{}_{1,m_I} \;, \quad  
\psi^{}_{0,0}= \phi^{a}_{}(z_1,z_2)\,\chi^{}_{0,0} \; . 
\eeq

As we have seen above, the action of the Hamiltonian on the direct sum of 
two sectors $\Gamma_{I=1} \oplus \Gamma_{I=0}$ is diagonal in each sector.
The eigenvalues of the operator $\Omega$ are given by (\ref{}).  
Thus, with respect to the coordinate part of the wave 
functions, there is a  ``bosonic'' sector for  $I=1$ and a ``fermionic'' 
sector for $I=0$, with  statistics parameters 
$-g/4$ and $3g/4$, and degeneracies 3 and 1, respectively. It
follows that the second virial
coefficient  can be expressed in terms of the 
second virial coefficients for Abelian anyons as 
\be
a_2^{\rm total} = \frac14 \left[3 a_2^{b}(-\frac{g}{4}) 
+ a_2^{f}(\frac{3g}{4}) \right] \:.
\label{a2-tot}\ee
This result is a generalization for $B \neq 0$ to the one obtained in 
\cite{Hagen}.
Remember that $0<g<2$: in the fermionic
sector,  $3g/4$ can fall outside the interval of
definition $[-1,1]$. In this case, the Abelian second virial coefficient
should be extended periodically, which means that  (\ref{a2-tot}) can still be
used but with $3g/4\to 3g/4-2$. As a consequence, a cusp appears in the 
second virial coefficient for $g=4/3$, as shown in Fig. 3.

In the vanishing magnetic field limit, the  second virial coefficient
interpolates between its end values for bosons and fermions
 with two internal 
degrees of freedom  when $g$ varies from
$g=0$ to $g=2$.

\begin{figure}
\centerline{\hbox{
   \epsfig{figure=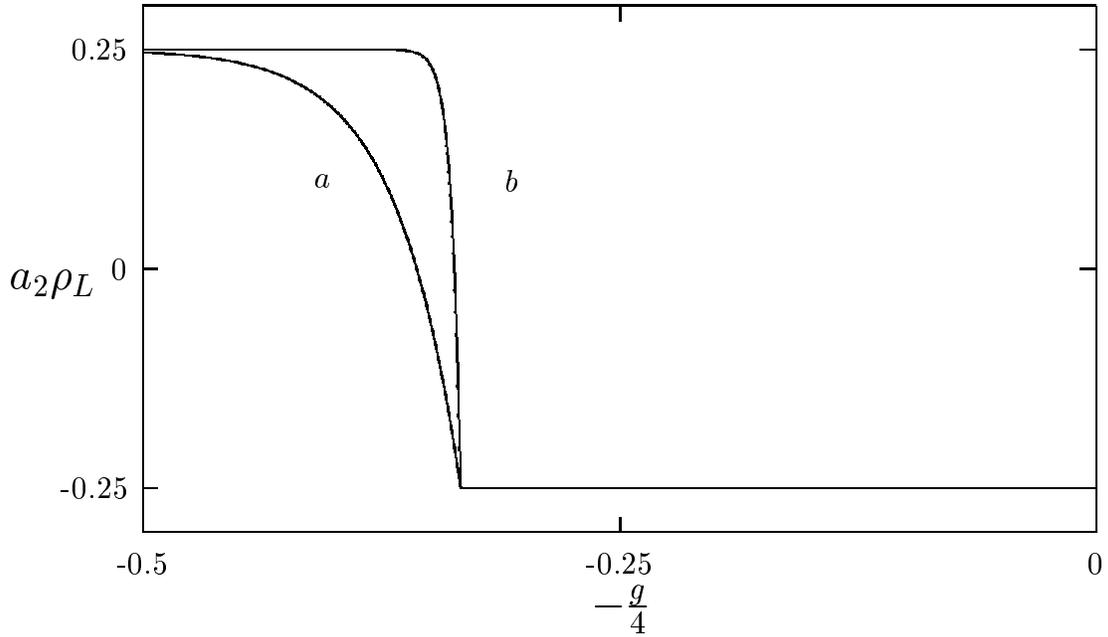,width=13cm}}
 }
 \caption{Second virial coefficient for the non-Abelian problem
as a function of the statistics parameter of the bosonic sector
$-\frac{g}{4}$ for (a) $x=10$ and (b) $x=70$.}
 \label{plothdr}
\end{figure}

In the strong magnetic field limit, 
the lowest Landau level  second virial coefficient reads  if $0<g<4/3$,
\be
a_2 (g) = - {1\over 4\rho_L}\ee
and if $4/3<g<2$,
\be
a_2 (g) = {1\over 4\rho_L}      \left( 1-2 e^{-2x \left( \frac{3g}{4} -1\right)} \right) \;.
\ee
It is close to a step function,
changing abruptly  near  $g= 4/3$
on an interval of width $\sim 1/(\beta \omega_c)$, from its bosonic end value
$-1/4\rho_L$ to  its fermionic end value $1/4\rho_L$.

\section{The thermodynamics for a  $N$-body non-Abelian problem in a 
strong magnetic field}
\label{N-body}

We now ask the question about a possible generalization of the $2$-body
analysis to the $N$-body case.

Let us first remind that in the Abelian case, from the lowest Landau
level spectrum
(\ref{},\ref{}) one can deduce \cite{Notre} an
equation of state which coincides with Haldane's statistics thermodynamics 
for a gas of particles whose $1$-body spectrum is reduced
to a single level of energy $E=\omega_c$.
The virial coefficients are
\be a_n= (-{1\over \rho_{\rm L}})^{n-1} {1\over 
n}\{(1+\alpha)^n-\alpha^n\}\ee
The critical filling 
$\nu_{\rm cr}=-1/\alpha$
where the pressure diverges describes a non degenerate ground state with all 
the $\ell_i$'s null. In the {\sl singular gauge}, 
\be 
\psi'=\prod_{i<j} z_{ij}^{-\alpha}\exp(-{\omega_{\rm c}\over 2} 
\sum_i^Nz_i\bar z_i) 
\ee
When $\alpha=-1$, one recovers a Vandermonde determinant
built from $1$-body Landau eigenstates. Incidentally, in the
Haldane statistics point of view, when $\alpha=-m$, 
the non degenerate ground state coincides with the Laughlin eigenstates
at the critical filling $\nu_{\rm cr}=1/m$.

\subsection{Lowest Landau level non-Abelian case}
\label{LLLnonAbelian}

For particles with isospin $1/2$,
there is a one-to-one correspondence between the total isospin of the system 
and the Young diagram for the isospin wave function \cite{LIII}. The 
corresponding Young diagrams may have at most two raws, of length 
$\frac{N}{2}+I$ and $\frac{N}{2} -I$ for total isospin $I$ and will be d
denoted as  
$[(\frac{N}{2}+I)(\frac{N}{2} -I)]$. 
Since the total wave function is symmetric under interchange of particles,
the symmetry properties of the isospin and 
coordinate wave functions are determined by the same Young 
diagram (see Appendix \ref{AppendixYoung}). 

The space of solutions of the $N$-body problem 
 is decomposed into a direct sum of 
sectors with given  values of the total isospin:
\be
\underbrace{\frac12 \otimes   \cdots  \otimes \frac12}_N =  
\Gamma_{I=\frac{N}{2}} \oplus 
\underbrace{ 
\Gamma_{I=\frac{N}{2}-1} \oplus   \cdots  \oplus \Gamma_{I=\frac{N}{2}-1}  
}_{ (N-1)!} 
 \oplus \cdots  \oplus 
\underbrace{ 
\Gamma_{I= \frac{N}{2} - \left[ \frac{N}{2} \right]} \oplus   \cdots  \oplus 
\Gamma_{I= \frac{N}{2} - \left[ \frac{N}{2} \right]}
}_{d_{N, \frac{N}{2} - \left[ \frac{N}{2}\right] }}
\label{decomposition}
\ee
The number of times the subspace with total isospin $I$ appears in the 
decomposition (\ref{decomposition}) is given by 
\be
d_{N,I} = d_{[(\frac{N}{2} +I)(\frac{N}{2} -I)]} = 
\frac{N!(2I+1)}{(\frac{N}{2} +I + 1)!(\frac{N}{2} - I)!}
\ee
which is also equal to the dimension 
of the representation $[(\frac12 N +I)(\frac12 N -I)] $.
 
The energy spectrum (\ref{E-NI}) can be represented in the form 
\be
  E_{N,I} =  
 (\omega_t - \omega_c) \Omega_{N,I} (g)
                +  E_{N,I}^{\rm bosons} (\ell_1, \ldots , \ell_N ), 
\label{decomp}\ee 
where 
 
\be
 E_{N,I}^{\rm bosons} (\ell_1, \ldots , \ell_N ) =
 N \omega_t + 
  (\omega_t - \omega_c) \sum_{j=1}^{N} \ell_j  
\label{E-NI-b}\ee
is (up to a constant) the spectrum of $N$ SU(2) bosons in a harmonic well 
of  frequency $ (\omega_t - \omega_c)$ 
in the isospin $I$ sector(s); the numbers $\ell_i$ are 
discussed in detail in the next subsection.   
This implies that  
 $N$-particle partition function is   
\be
Z_{N} = 
\sum_{I= \frac{N}{2} - \left[ \frac{N}{2} \right]}^{ \frac{N}{2} } 
e^{-\beta (\omega_t - \omega_c)\, \Omega_{N,I} (g)} \, 
 Z_{N,I}^{\rm bosons}
\label{ZN-total}\ee
where $ Z_{N,I}^{\rm bosons}$ is the partition function 
of SU(2)-bosons determined by the spectrum (\ref{E-NI-b}).
 
For $g=0$, (\ref{ZN-total}) reduces to 
\be
 Z_{N}= 
       \sum_{I= \frac{N}{2} - \left[ \frac{N}{2} \right]}^{ \frac{N}{2}} 
 Z_{N,I}^{\rm bosons} = 
\sum_{k=0}^{N} {Z}_{N-k}^b {Z}_k^b \; ,
\label{ZN-total-b}\ee
where ${Z}_{k}^{ b}$ is the usual $N$-particle partition function for 
bosons without internal  degrees of freedom.
The latter equality in (\ref{ZN-total-b})
follows from the fact that the grand partition function
for isospin $1/2$ bosons is the square of the grand partition
function of bosons without internal degrees of freedom.
 
The partition functions for SU(2) bosons 
$ Z_{N,I}^{\rm bosons} $ are discussed in 
the next subsection, where these are  related to 
partition functions associated with Young diagrams 
(see (\ref{bosons-Young})).

\subsection{Partition functions associated with Young diagrams}
\label{Z_Y}

We start with a system of identical particles without internal degrees of
freedom in a harmonic well. 
The single-particle energy spectrum is  
\be
\e_{\ell} = (\omega_t-\omega_c)  \ell 
\label{epsilon-l}\ee
with $\ell = 0,1,2, \ldots $, where we have left aside the constant
 ground state energy shift $\omega_t$ which has no effect on the thermodynamics.

The $2$-body wave function of arbitrary permutation symmetry can be represented 
as a linear combination of symmetric and   
antisymmetric wave functions.
This corresponds to decomposition of the set of all possible values of 
the quantum numbers $ ( \ell_1, \ell_2 )$, 
$\ell_1, \ell_2 =0,1,2, \ldots $ for two particles 
into 
\be
\ell_1 \leq \ell_2   \quad \quad {\rm and } \quad \quad
\ell_1 < \ell_2
\label{ineq2}\ee
Correspondingly, the Boltzmann partition function is
the  sum of the partition functions for bosons and fermions ($q \equiv e^{-\beta
(\omega_t-\omega_c)}$)
\be
 Z_{2}^{\rm Boltz}\equiv \frac{1}{(1-q)^2} 
= Z_{2}^{\rm b} + Z_{2}^{\rm f} 
= Z_{\jd} + Z_{\juu}
\label{Boltz2}\ee
where 
\be
 Z_{\jd} = \frac{1}{(1-q)(1-q^2)} \; ,\quad 
 Z_{\juu} = \frac{q}{(1-q)(1-q^2)} \;.
\label{Z2-bf}\ee

In the $3$-body case, a wave function of an arbitrary permutation 
symmetry can be decomposed into a sum of wave functions with
three  types of symmetry, bosonic, 
fermionic and mixed. Note that 
there are two (identical) representations 
of the mixed symmetry and that  these representations are
two-dimensional\footnote{The above decomposition of the wave function is a  
decomposition of the {\em regular} representation of 
$S_3$ into irreducible representations.}.
To discuss decomposition of  all possible 
values of the quantum numbers   $ ( \ell_1, \ell_2, \ell_3 )$, for three 
particles, one should consider standard Young { \it tableaux} 
labeling all possible symmetries of wave functions (see Appendix 
\ref{AppendixYoung}).

With each  standard  Young tableau we associate a
 $\ell$-Young tableau
which is obtained by the change $i \to \ell_i $ for all the numbers
inside the standard  Young tableau. In the $3$-body case, we then obtain 
four  $\ell$-Young tableaux 

\begin{figure}[h]
\begin{center}
\begin{picture}(20,10)
\put(0,0){\framebox(5,5){$\ell_1$}}
\put(5,0){\framebox(5,5){$\ell_2$}}
\put(10,0){\framebox(5,5){$\ell_3$}}
\put(17,0){,}
\end{picture}
\begin{picture}(15,10)
\put(0,2.5){\framebox(5,5){$\ell_1$}}
\put(5,2.5){\framebox(5,5){$\ell_2$}}
\put(0,-2.5){\framebox(5,5){$\ell_3$}}
\put(12,0){,}
\end{picture}
\begin{picture}(15,10)
\put(0,2.5){\framebox(5,5){$\ell_1$}}
\put(5,2.5){\framebox(5,5){$\ell_3$}}
\put(0,-2.5){\framebox(5,5){$\ell_2$}}
\put(12,0){,}
\end{picture}
\begin{picture}(10,10)
{and}
\end{picture}
\begin{picture}(15,10)
\put(0,-5){\framebox(5,5){$\ell_3$}}
\put(0,0){\framebox(5,5){$\ell_2$}}
\put(0,5){\framebox(5,5){$\ell_1$}}
\put(7,0){.}
\end{picture}
\end{center}
\caption{$\ell$-Young tableaux for three particles}
\label{l-YTabl3}
\end{figure}

\noindent
The second and third tableaux are associated with the 
two standard Young tableaux for the states of mixed symmetry.

All possible values of $\ell_1$, $ \ell_2$, 
and $\ell_3$ can be arranged as
\bea
\ell_1 \leq \ell_2 \leq \ell_3 \;, \label{ineq3-a}   \\
\ell_1 \leq \ell_2  < \ell_3 \;, \label{ineq3-b}  \\    
\ell_1 < \ell_2 \leq \ell_3 \;,  \label{ineq3-c} \\
\ell_1 <  \ell_2 < \ell_3 \;,  \label{ineq3-d}
\eea 
where (\ref{ineq3-a}) and (\ref{ineq3-d}),
corresponding  to bosons and fermions, respectively lead to the partition
functions 
\bea
 Z_{\jt} &=& \frac{1}{(1-q)(1-q^2)(1-q^3)} \; , \nonumber \\
 Z_{\juuu} &=& \frac{q^3}{(1-q)(1-q^2)(1-q^3)} \; .
\label{Z3-bf}\eea
Straightforward calculations show that 
(\ref{ineq3-b}) and 
(\ref{ineq3-c}) lead to the partition functions 
$2q/[(1-q)(1-q^2)(1-q^3)]$ and 
$2q^2/[(1-q)(1-q^2)(1-q^3)]$, respectively. 
We associate these contributions with the second and third $\ell$-Young 
tableaux, respectively, in Fig. \ref{l-YTabl3}.  
Thus the partition function of
the Young diagram of  mixed symmetry is 
\be
 Z_{\jdu} = \frac{2(q + q^2)}{(1-q)(1-q^2)(1-q^3)} \; .
\label{Z2mixed}\ee
The Boltzmann partition function rewrites as  
\be
 Z_{3}^{\rm Boltz}\equiv \frac{1}{(1-q)^3} =
  Z_{\jt} +  Z_{\jdu} +  Z_{\juuu}
\label{Boltz3}\ee
as it should.

We are now in position to propose
general rules to write down partition functions
associated with  Young diagrams in the $N$-body case. To this aim, we represent the   
partition function associated with a Young diagram $Y_N$ with $N$ boxes as 
\be
Z_{Y_N}= d_{Y_N}\frac{P_{Y_N}(q)}{(q)_N} \;, \quad
(q)_n \equiv \prod_{k=1}^n(1-q^k)\; . 
\label{ZYN}\ee
where $d_{Y_N}$ is the dimension of the irreducible representation 
of the symmetric group $S_N$ corresponding to the Young diagram 
$Y_N$ (equal to the number of standard Young tableaux for $Y_N$),
and $P_{Y_N}$ is a polynomial in $q$. 
Note that the partition functions for $N$ Abelian 
bosons and fermions read 
$Z_{N}^{b} = {1}/[{(q)_N}] $
and 
$ Z_{N}^{f} = {q^{N(N-1)}}/[{(q)_N}] $, respectively.

To evaluate
$P_{Y_N}(q)$, to each $\ell$-Young tableau is associated a {\it chain}  of inequalities
for the  integer numbers
$\{ \ell_1\le \ell_2\le \ldots\le \ell_N \}$ in the following way:
If  the row containing
the box with $\ell_{j+1}$ is located below the raw containing the box
with $\ell_{j}$,
then $\ell_j<\ell_{j+1}$; otherwise,  $\ell_j\le\ell_{j+1}$.
Let $s_1, s_2, \ldots $ be the positions of the signs `$<$' in this chain
counted from the right.
Then the contribution to the polynomial $P_{Y_N}$
from a given $\ell$-Young tableau is  $q^{\sum_j s_j}$.
 Finally,  $P_{Y_N}$, associated with a given Young diagram,
is the sum of the contributions from
all $\ell$-Young tableaux for this diagram.

For instance, for (\ref{ineq3-b}),
$s_1=1$,  and the contribution to  $P_{[21]}$
is  $q$ while
for  (\ref{ineq3-c}), $s_1=2$, and its contribution
is $q^2$.
For the $4$ and $5$-body cases, explicit expressions for $P_{Y_N}$ are 
given in Appendix \ref{AppendixYoung}. 

Notice a relation between the partition functions for 
$Y_N$ and its  conjugated Young diagram $\bar{Y_N}$: if
$P_{Y_N} = \sum_{k=1}^{k_{\rm max}} c_{Y_N}(k) q^k $
then $ P_{\bar{Y_N}} = \sum_{k=1}^{k_{\rm max}} c_{Y_N}(k) q^{k_{\rm max}-k}$.

As a consistency check,
all possible partition functions in the $N$-body 
case should sum to the $N$-body Boltzmann partition function, 
generalizing (\ref{Boltz2}) and (\ref{Boltz3})  
\be
Z_{N}^{\rm Boltz}= \frac{1}{(1-q)^N}= 
\sum_{Y_N} Z_{Y_N}  \;.
\label{Boltz-constr}\ee
Using the partition functions above
we checked using {\it Mathematica}, up to the $6$-body case, that 
(\ref{Boltz-constr}) is indeed valid.  

Now we consider SU(2) bosons. 
The decomposition of the space of solution of the $N$-body problem 
into sectors with different isospins   
was discussed in the previous section. 
Taking into account the multiplicity $2I+1$ for isospin $I$ states, 
the partition function for SU(2) bosons in the isospin $I$ sector(s)
can be related to the partition function associated with the Young diagram
corresponding to isospin $I$ as follows 
\be
Z_{N,I}^{\rm bosons} = (2I+1) Z_{[(\frac{N}{2} +I)(\frac{N}{2} -I)]} \;.
\label{bosons-Young}\ee
Note that $ Z_{[(\frac{N}{2} +I)(\frac{N}{2} -I)]}$ receives 
contributions from  $d_{[(\frac{N}{2} +I)(\frac{N}{2} -I)]}$
$\ell$-Young tableaux. This corresponds to taking into
account all $d_{[(\frac{N}{2} +I)(\frac{N}{2} -I)]}$ sectors of  given $I$ in 
the decomposition of the space of
solutions of $N$ SU(2) bosons in calculating their
partition functions.     
One can check, using  (\ref{ZYN}) and the expressions for the 
polynomials  given in this Section and 
in Appendix \ref{AppendixPolynomials}, 
that (\ref{bosons-Young}) 
leads to the correct expression for the total 
partition functions for $N$ bosons (\ref{ZN-total-b}).

To conclude this section, we  comment on the relation between 
the above partition functions and the Schur functions.
Schur functions naturally arise for the system of non interacting
particles whose $1$-body spectrum is represented 
by a finite set of levels $\e_1, \e_2, \ldots , \e_M$ \cite{schur}.
One defines $x_1 = e^{-\beta \e_1}$, $\ldots$, 
 $x_M = e^{-\beta \e_M}$.
Then the Boltzmann partition function
can be represented as a sum over 
all irreducible representations $\lambda $ of $S_N$
 \bea
Z_N^{\rm Boltz}(x_1,\ldots,   x_M)  & \equiv&   (x_1+\cdots   +   x_M)^N 
\nonumber \\
&  =&
\sum_{{\lambda\atop |\lambda|=N}} d(\lambda)  S_\lambda  (x_1,\ldots,
x_M)\,\,\,\,,
\label{schur}
\eea
where $S_\lambda(x_1,\ldots,x_M)$ are
the Schur functions
associated with the irreducible representations  $\lambda $ of
the symmetric group $S_N$.
The case  discussed above corresponds to $\e_k = (k-1) (\omega_t-\omega_c) $ and
 $M \to \infty $. We thus observe that the partition functions discussed 
in this section are a generalization of the Schur functions  
to an infinite number of arguments corresponding to
an infinite set of equally spaced $1$-body energy levels. 
These functions can be used in studies of the statistical mechanics of 
systems with parastatistics \cite{schur,PolychronakosNPB}

\subsection{On the virial expansion  for   
non-Abelian Chern-Simons particles in a strong magnetic field}
\label{virial-N-LLL}

Unlike the Abelian case, an explicit
expression for the $N$-body partition functions is not available.
Nevertheless, it is possible to construct iteratively
the cluster  coefficients $b_k^{\omega}$ of order $k$ for any $k\leq N$.
The cluster coefficients in 
the thermodynamic limit are obtained as
\beq
b_k= V k \rho_L \beta \lim_{\omega\to 0} (\omega_t-\omega_c)  b_k^{\omega}
\label{bkV}
\ee

Using the partition functions (\ref{ZN-total}) 
with (\ref{bosons-Young}), we find 
the cluster coefficients of lowest orders 
\bea
b_2^{\omega}&=&(-1 + \Omega_{2,0} + 3 \Omega_{2,1}) \frac{1}{\beta
(\omega_t-\omega_c)} + ...
\\
b_3^{\omega}&=&\left(\Omega_{2,0} + 3 \Omega_{2,1} -
\frac{2}{3}( \Omega_{3,\frac{1}{2}} + \Omega_{3,\frac{3}{2}}) \right) 
\frac{1}{\beta^2 (\omega_t-\omega_c)^2}
+ \nonumber \\
& &
\left(\frac{2}{9} + \Omega_{2,0}(1-\frac{1}{2}\Omega_{2,0})
+6 \Omega_{2,1} ( 1-\frac{1}{4}\Omega_{2,1}) \right. -
\nonumber \\ & &
\left. \Omega_{3,\frac{1}{2}}(1-\frac{1}{3}\Omega_{3,\frac{1}{2}}) -
2 \Omega_{3,\frac{3}{2}} (1- \frac{1}{6}\Omega_{3,\frac{3}{2}}) \right ) 
\frac{1}{\beta (\omega_t-\omega_c)}+\ldots
\eea
In the thermodynamic limit,
$ b_k^{\omega}$ should
behave as $(\omega_t-\omega_c)^{-1}$, i.e.  for small $\omega$
$b_{k}$ should be 
proportional to the volume. It is easy indeed
to verify that  $\Omega_{N,I}$ in
(\ref{Omega1})
is such that the correct thermodynamic limit is obtained.

Moreover,  $\Omega_{N,I}$ is such that, in the cluster expansion,
the leading term  $(\omega_t-\omega_c)^{-1}$ {\it does not
depend on $g$} what we checked up to the fifth virial coefficient.
This implies that the cluster
coefficients (and consequently the virial
coefficients) do not depend on the statistics parameter.
It would be interesting to understand a possible symmetry 
underlying this cancellation.

The above results agree  with  the exact second virial
coefficient analysis if
$g$ is sufficiently small, $|g| \leq 4/3$.
However, at $g=4/3$ there is  a jump in the second
virial coefficient.
It follows that
the $N$-body spectrum (\ref{E-NI}) {should not be considered } 
as the correct spectrum
for arbitrary $g$. As we have seen in the
$2$-body  case, a shift in the statistics parameter
had to be done, which in turn lead to the jump in the second virial coefficient. 
We have not found a generalization of this 
procedure to the  $N$-body case.

In analogy with the Abelian case, we can argue
that the spectrum of anyons is indeed given
by (\ref{E-NI}) if $\Omega_{N,I}$
are less than one. Under this assumption,
it is easy to show that the cluster coefficients
$b_k$ are independent of $g$ when

\be
 0 \leq g \leq  \frac{8}{k(k-1)} \;, \quad  \quad k \geq 3 \;. 
\label{g-range}
\ee  
In this range the cluster coefficients are the same as those of 
SU(2) bosons. The behavior of the cluster (and virial) coefficients 
in the entire interval of the statistics parameter is yet an open question. 

\section{Concluding remarks}
\label{conclusion}

We have developed a formalism  to study the thermodynamics of
non-Abelian Chern-Simons particles 
in the lowest Landau level of an external magnetic field. 
We expect that  the thermodynamics of non-Abelian Chern-Simons 
particles described by 
Hamiltonians  of the type 
(\ref{nabe1},\ref{nabe2}) and  (\ref{25},\ref{26})  
with other groups,  as well as  other irreducible 
representations, can be analyzed using
the same ideas with
minor modifications.
The hope is that under  some appropriate choices of the non-Abelian symmetry 
one can obtain non-trivial thermodynamics in the lowest Landau level.

In this respect, 
an attractive possibility  would be to use non-Abelian symmetries  
proposed recently in Ginzburg-Landau  Chern-Simons 
theories for non-Abelian quantum Hall states \cite{GLCSnonAbelian}. 
An argument in favor of this choice is the new 
exclusion statistics thermodynamics  
found for non-Abelian quantum Hall states 
quasiparticles on the edge \cite{Schoutens-pfaffian}
and for their bulk counterparts \cite{SIunpublished}. 
We plan to address this issue in future publications.

Another  observation is
 that the same  form of the spectrum, namely a  bosonic $N$-body spectrum 
plus a  linear term in the statistics parameter, arises 
in one-dimensional integrable models with inverse square interaction \cite{Poly}.
In those models, the coupling to the Chern-Simons field is 
replaced by the coupling of the inverse square interaction. 
The  spectrum is then  valid in the entire interval of definition 
of the coupling  parameter. It follows that    
 that the cluster and virial coefficients do not depend on 
the coupling parameter in the entire interval of definition of the 
interaction parameter. 
It would certainly be rewarding to find a symmetry principle
underlying this cancellation. 

\acknowledgments 
We would like to thank A. P. Polychronakos for interesting discussions.
G.L. is supported by an European Union grant ERBFMBICT 961226. 

\appendix
\section{Young diagrams and irreducible representations}
\label{AppendixYoung}

For completeness we include in this appendix a brief
description of the relation between Young diagrams and
representations of the permutation group of N particles $S_N$. 
In particular we show how wave functions of 
arbitrary permutation symmetry can be constructed using Young operators.

\begin{figure}[h]
\begin{center}
\begin{picture}(15,10)
\put(0,0){\framebox(5,5){$$}}
\put(5,0){\framebox(5,5){$$}}
\end{picture}
\begin{picture}(15,10)
\put(0,2.5){\framebox(5,5){$$}}
\put(0,-2.5){\framebox(5,5){$$}}
\end{picture}
\begin{picture}(20,10)
\put(0,0){\framebox(5,5){$$}}
\put(5,0){\framebox(5,5){$$}}
\put(10,0){\framebox(5,5){$$}} 
\put(17,0){}
\end{picture}
\begin{picture}(15,10)
\put(0,2.5){\framebox(5,5){$$}}
\put(5,2.5){\framebox(5,5){$$}}
\put(0,-2.5){\framebox(5,5){$$}}
\put(12,0){}
\end{picture}
\begin{picture}(15,10)
\put(0,-5){\framebox(5,5){$$}}
\put(0,0){\framebox(5,5){$$}}
\put(0,5){\framebox(5,5){$$}}
\put(7,0){}
\end{picture}
\caption{Young diagrams for two and three particles}
\end{center}
\label{YDiagr23}
\end{figure}
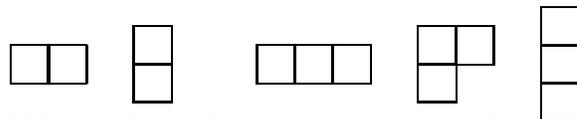

The irreducible representations
of  $S_N$ can  be associated with  the partitions
of $N$ in positive integers $\lambda_i$,
\beq
\lambda_1 + \lambda_2 + ... + \lambda_m = N  \; , \quad 
\lambda_1 \geq \lambda_2 ...\geq \lambda_m \;.
\label{part}
\eeq
The different partitions (\ref{part}) can be depicted graphically
by means of  {\it Young diagrams} in which each number   $\lambda_i$
is represented by a row of  $\lambda_i$ cells. 
The partitions (\ref{part})
are usually denoted by $[\lambda]=
[\lambda_1 \lambda_2...\lambda_m]$
where the power is used to indicate the repeated appearance of
the same integer. Thus for two particles we have,
$[\lambda]=[2]$ and $[\lambda]=[1^2]$  
associated with the first two  diagrams in Fig \ref{YDiagr23}, while for 
three particles we have
$[\lambda]=[3]$, $[\lambda]=[21]$, and $[\lambda]=[1^3]$ 
corresponding to the last three diagrams in Fig.\ref{YDiagr23}.
 
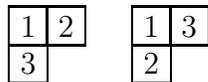
\begin{figure}[h]
\begin{center}
\begin{picture}(15,10)
\put(0,2.5){\framebox(5,5){$1$}}
\put(5,2.5){\framebox(5,5){$2$}}
\put(0,-2.5){\framebox(5,5){$3$}}
\put(12,0){}
\end{picture}
\begin{picture}(15,10)
\put(0,2.5){\framebox(5,5){$1$}}
\put(5,2.5){\framebox(5,5){$3$}}
\put(0,-2.5){\framebox(5,5){$2$}}
\put(12,0){}
\end{picture}
\end{center}
\caption{Standard Young tableaux for the { [21]} representation}
\label{yd2}
\end{figure}

The number of inequivalent irreducible representations
of $S_N$ is equal to the number of different Young diagrams.
In our examples, $S_2$ and $S_3$ have two and three inequivalent
representations respectively. 
Using Young diagrams it is
also possible to determine the dimensions of such representations
as well as to construct explicitly representations 
$\Gamma^{[\lambda]}(P)$
of the elements $P \in S_N$ and
the basis vectors. To do that, 
a {\it standard} Young tablueax is defined in which
the numbers from $1$ to $N$ are distributed among the boxes of
a Young tableau in such a way that they increase from left
to right along the same row and from top to bottom along
the  the same columns. Then 
the dimension of the representation
is equal to the number of standard Young tableaux.
For example, there is only one standard Young tableau
for each of the diagrams in Fig \ref{YDiagr23}, except for ${[21]}$,
and so they correspond to one-dimensional representations
of the permutation group. The diagram  ${ [21]}$ has 
two standard Young tableaux shown in Fig \ref{yd2},
corresponding to a two-dimensional representation.

\noindent
Since any element of $S_N$ can be written as a 
the product of $N-1$ transpositions  $P_{i-1,i}$, it is sufficient   
to specify the matrices 
$\Gamma^{[\lambda]}(P_{i-1 i})$ representing these transpositions. 
The following rules define the {\it Young-Yamanouchi standard} 
orthogonal representation \cite{Kaplan}
where superscripts $r , t$ run over all standard Young tableaux
for the Young diagram $[\lambda]$.

\bigskip
{\it Diagonal elements}

\begin{enumerate}
\item $\Gamma^{[\lambda]}_{rr}(P_{i-1 i }) = 1$ if in the tableau
r, $i$ and and $i-1$ are on the same row.

\item  $\Gamma^{[\lambda]}_{rr}(P_{i-1 i }) = -1$ if in the tableau
r, $i$ and and $i-1$ are on the same column.

\item $\Gamma^{[\lambda]}_{rr}(P_{i-1 i }) = \pm \frac{1}{d}$, otherwise.
 Here $d$ is the number
of vertical and horizontal steps that  one must take to move from
$i-1$ to $i$, and the upper (lower) signs applies when the row containing
the number $i$ is above (below)  the one containing $i-1$.
\end{enumerate}
{\it Non-diagonal elements}
\begin{enumerate}
\item $\Gamma^{[\lambda]}_{rt}(P_{i-1 i }) = (1-\frac{1}{d^2})^{\frac{1}{2}} $, if
the tableaux $r$ and $t$ differ only by a permutation of numbers $i$ and $i-1$
and where $d$ is defined as above.
\end{enumerate}
\noindent
For instance, for the two-dimensional representation ${ [21]}$
we only need to find $\Gamma^{[21]}(P_{12})$ and  $\Gamma^{[21]}(P_{23})$
(as $P_{13}=P_{12}P_{23}P_{12}$ , $P_{123}=P_{12}P_{23}$ and $P_{132}=P^{-1}_{123}$).
The application of the rules above gives
\beq
\Gamma^{[21]}(P_{12})  =  \left(
\begin{array}{cc}
1 & 0 \\
0 & -1 
\end{array}
\right)  \;,
\;\;\;\;\;\;\;\;\;\;
\Gamma^{[21]}(P_{23}) = \left(
\begin{array}{cc}
-\frac{1}{2} & \frac{\sqrt{3}}{2} \\
\frac{\sqrt{3}}{2} & \frac{1}{2}
\end{array}
\right) \;.
\eeq

To construct the basis  for the  representation $[\lambda]$, 
$N!$ { \it Young operators} are introduced as 
\beq
\omega^{[\lambda]}_{rt}= (\frac{f_{\lambda}}{N})^{\frac{1}{2}} \sum_{P} 
 \Gamma^{[\lambda]}_{rt}(P) P    \; .
\eeq
where $f_{\lambda}$ is the dimension of the representation.
These $f_{\lambda}$ operators with a fixed second index 
transform into each other under permutations and 
form a basis for the irreducible representation $\Gamma^{[\lambda]}$.

A basis more convenient in physical applications 
can be obtained by applying the
Young operators to a product of ``one-particle wave functions" in
the following way. Consider  a set of $N$ orthogonal
functions $\psi_a(i)$, where $i$ stands for the
sets of variables on which the $\psi_a$ depend,  and 
form the product wave function $\Psi_0$,
\beq
\Psi_0= \psi_1(1)\psi_2(2) \cdots \psi_N(N)   \;.
\eeq
One can then obtain a basis for the orthogonal representations
by applying the Young operators to this function
\beq
\Psi^{[\lambda]}_{rt}= \omega^{[\lambda]}_{rt} \Psi_0  \; .
\eeq
where $P$ is understood as acting on the $arguments$ of $\Psi_a$.
Then, the functions $\Psi^{[\lambda]}_{rt}$ with a fixed {\it second}
index transform into each other under permutation and form a basis
for a irreducible representation of the permutation group. The
Young tableau $r$ that corresponds to the first index,
enumerates the different wave functions in a given basis
and characterizes the symmetry under the interchange of arguments.
The second index $t$ enumerates the different basis for 
$\Gamma^{[\lambda]}$ and it can be shown that it characterizes
the symmetry of the wave function under the interchange
of the function $\psi_a$.

One can apply the above general discussion to the case of the coordinate wave
function of a 3-body system.
Consider a coordinate wave function 
$\phi(123) \equiv \phi(\bfr_1,\bfr_2,\bfr_3)$
(eventually the product of single particle wave functions,
$  \phi(\bfr_1,\bfr_2,\bfr_3) = \phi_{\alpha}(\bfr_1) \phi_{\beta}(\bfr_2)
\phi_{\gamma}(\bfr_3)$). 
Then in addition to the totally symmetric and totally antisymmetric 
wave functions,
\bea
\phi^{[3]}&=&\frac{1}{\sqrt{6}}
\Bigl( \phi(123) + \phi(123) + \phi(132) + \phi(321) +  
\phi(231) +  \phi(312) \Bigr)  \;, \nonumber \\
\phi^{[1^3]}&=&\frac{1}{\sqrt{6}}
\Bigl( \phi(123) - \phi(123) - \phi(132) - \phi(321) +  
\phi(231) +  \phi(312) \Bigr) \;, 
\label{phi_symm}
\eea
one obtains for the mixed symmetry wave functions 
for the $two$ 2-dimensional representation ${ [21]}$
\bea
\phi^{{ [21]}}_{11}&=& \frac{1}{\sqrt{12}} \Bigl( 
2 \phi(123) + 2 \phi(213) - \phi(132)      
- \phi(321) - \phi(231) - \phi(312) \Bigr)  \;, \nonumber \\
\phi^{[21]}_{21} &=& \frac{1}{2} \Bigl(
\phi(132)  -   \phi(321) - \phi(231) + \phi(312) \Bigr) \;,
\label{phi_mixed1}
\eea
\bea
\phi^{[21]}_{22}&=& \frac{1}{\sqrt{12}} \Bigl( 
2 \phi(123) - 2 \phi(213) + \phi(132)      
+ \phi(321) - \phi(231) - \phi(312) \Bigr) \;, \nonumber \\
\phi^{[21]}_{12} &= &\frac{1}{2} \Bigl(
\phi(132)  - \phi(321) + \phi(231) - \phi(312) \Bigr) \;,
\label{phi_mixed2}
\eea

For a system of bosons with two internal degrees of freedom,
the total wave functions for a given total isospin $I$  
(symmetric under interchange of 
particles) can be written as 
\be
\psi_{\alpha \beta}= \sum_{\gamma} \phi^{[\lambda]}_{\gamma \alpha} 
\chi^{[\lambda]}_{\gamma \beta} \;,
\label{totalwf}
\ee
where  $ \phi^{[\lambda]}_{\gamma \alpha}$
and $\chi^{[\lambda]}_{\gamma \beta}$ are the coordinate and isospin wave 
functions of the permutation symmetry 
$[\lambda] = [(N/2+I)(N/2-I)]$.

Again in the 3-body case, the eight-dimensional internal space can
be written as the sum of a $I=3/2$ four-dimensional space,
associated with wave functions which are completely symmetric
under the interchange of isospin indices, plus two $I=1/2$ two-dimensional
spaces associated with states of mixed symmetry.
The basis of the $I=3/2$ subspace is  
therefore
\bea
\chi_{\frac{3}{2},\frac{3}{2}}  &=& |+ + + > \;, \nonumber \\
\chi_{\frac{3}{2},\frac{1}{2}} &=& 
\frac{1}{\sqrt{3}} ( |+ + -> + | + - + > + | - +  + > ) \;, \nonumber \\
\chi_{\frac{3}{2},-\frac{1}{2}}  &=& 
\frac{1}{\sqrt{3}} ( |- - + > + | - + - > + | + -  - > ) \;, \nonumber \\
\chi_{\frac{3}{2},-\frac{3}{2}} &=&  |- - - > \;.
\eea
For $I=1/2$ there are {\it two} two-dimensional representations, one 
generated by 
\bea
\chi^{(1)}_{\frac{1}{2},\frac{1}{2}} &=&  
\frac{1}{\sqrt{6}} ( 2 |+ + -> - | + - + > + | - +  + > ) \;, \nonumber \\
\chi^{(1)}_{\frac{1}{2},-\frac{1}{2}} &=& 
\frac{1}{\sqrt{6}} ( 2 |- - + > - | - + - > + | + -  - > ) 
\eea
and the other by
\bea
\chi^{(2)}_{1,\frac{1}{2}}  &=&  
\frac{1}{\sqrt{2}} ( |+ - +> - | - +  + > ) \;, \nonumber \\ 
\chi^{(2)}_{1,-\frac{1}{2}}  &=&  
\frac{1}{\sqrt{2}} ( |- + -> - | + -  - > )  \; .
\eea
The superscript in these formulas  is related to the permutation symmetry.
For a fixed total isospin projection, say $I_z=1/2$, 
$\chi^{(1)}_{\frac{1}{2},\frac{1}{2}}$
and $\chi^{(2)}_{\frac{1}{2},\frac{1}{2}}$ form a 
basis for a two-dimensional representation of $S_3$. 
More precisely, they correspond to 
$\Psi^{[21]}_{1a}$ and $\Psi^{[21]}_{2a}$, 
with $a=1$ or $a=2$ having taken
 $\Psi_0(1,2,3)=|+-+>$. Notice that since there are only two available states, 
$+$ and $-$, $a=1$ and $a=2$ give the {\it same} results so that  
only two independent wave functions can be formed. 
The same argument applies to the $I_z=-1/2$ sector starting
from $\Psi_0 (1,2,3)=|-+->$.

According to (\ref{totalwf}), the total wave functions for the  system of
bosons are obtained by multiplying coordinate and isospin wave functions
of appropriate symmetries. For the sector $I=3/2$, one has four 
wave functions    
\be
\psi_{\frac{3}{2},m_I}= \phi_{\frac{3}{2},m_I}(z_1,z_2,z_3) 
                         \chi_{\frac{3}{2},m_I} \;,
\ee
where $\phi_{\frac{3}{2},m_I}(z_1,z_2,z_3)$ is a totally symmetric 
wave function (\ref{phi_symm}). 
For two representations in the isospin $I=1/2$ (mixed symmetry) sector, 
we obtain four basis wave functions    
\be
\psi^{(1)}_{\frac{1}{2} ,m_I} = \phi^{(11)}_{\frac{1}{2},m_I} 
 \chi^{(1)}_{\frac{1}{2},m_I}
  + \phi^{(21)}_{\frac{1}{2},m_I} \chi^{(2)}_{\frac{1}{2},m_I} \;,
\ee
\be
\psi^{(2)}_{\frac{1}{2},m_I} =  
\phi^{(12)}_{\frac{1}{2},m_I} \chi^{(1)}_{\frac{1}{2},m_I}
           + \phi^{(22)}_{\frac{1}{2},m_I} \chi^{(2)}_{\frac{1}{2},m_I} \;,
\ee
where  $\phi^{(ab)}_{1,m_I}$ with $a,b=1,2$ are the same as  
 $\phi^{[21]}_{ab}$ in (\ref{phi_mixed1}-\ref{phi_mixed2}).

\section{Polynomials $P_{Y_n}(q)$  for four and five particles}
\label{AppendixPolynomials}

Following the rules of subsection \ref{Z_Y}, 
one obtains the following expressions for 
the polynomials  $P_{Y_n}(q)$ in Eq.~(\ref{ZYN}) for four particles
\bea   
P_{\jq} (q)  &=&  1   \ \nonumber \\   
P_{\jtu} (q) &=&  q + q^2 + q^3   \nonumber \\    
P_{\jdd}  &=&  q^2 + q^4   \nonumber \\ 
 P_{\jduu}  &=& q^3 + q^4 + q^5    \nonumber \\ 
P_{\juuuu}  &=&  q^6  
\label{P4}\eea
and for five particles
\bea   
P_{[5]} (q)  &=&  1    \nonumber \\   
P_{[41]} (q) &=&  q + q^2 + q^3 + q^4    \nonumber \\    
P_{[32]} &=&  q^2 + q^3 + q^4 + q^5 + q^6   \nonumber \\ 
P_{[3 1^2]}  &=& q^3 + q^4 + 2 q^5 + q^6 + q^7  \nonumber \\ 
 P_{[2^2 1]}  &=& q^4 + q^5 + q^6 + q^7 + q^8   \nonumber \\ 
 P_{[2 1^3]}  &=&  q^6 + q^7 + q^8 + q^9     \nonumber \\ 
P_{[1^5]}  &=&  q^{10}  \;.
\label{P5}\eea
Note that the sum of all the coefficients of $P_{Y_n}(q)$ is equal 
to the dimension 
$d_{Y_N}$ of the associated respresentation of $S_N$.


\begin{thebibliography}{40}


\bibitem{Anyon}  
J. M. Leinaas and J. Myrheim, Nuovo Cimento {\bf B37}, 1 (1977); G. A.
Goldin, R. Menikoff and D. H. Sharp, J. Math. Phys. {\bf 22}, 1664 (1981);
F. Wilczek,  Phys. Rev. Lett. {\bf 48}, 1144 (1982); {\bf 49}, 957 (1982)


\bibitem{Calogero-Sutherland} 
F. Calogero, J. Math. Phys. {\bf 10}, 2191 (1969); {\bf 12}, 419 (1971);
B. Sutherland,  Phys. Rev. {\bf A4}, 2019 (1971); {\bf A5}, 1372 (1972)

\bibitem{Haldane} 
F. D. M. Haldane, Phys. Rev. Lett. {\bf 67}, 937 (1991)

\bibitem{Notre} 
A. Dasni\`eres de Veigy and S. Ouvry, Phys. Rev. Lett. {\bf 72}, 600 (1994)

\bibitem{ES} 
S. B. Isakov,
\journal Mod. Phys. Lett. B, 8, 319, 1994; Y.-S. Wu, \journal Phys. Rev.
Lett., 73, 922, 1994;
A. K. Rajakopal, \journal Phys. Rev. Lett., 74, 1048, 1995.

\bibitem{ES-integrable} 
S. B. Isakov, \journal Int. J. Mod. Phys. A, 9, 2563, 1994; 
D. Bernard and Y.-S. Wu, cond-mat/9404025;  
Z. N. C. Ha, \journal Phys. Rev. Lett., 73, 1574, 1994; 
M. V. N. Murthy and R. Shankar, \journal Phys. Rev. Lett., 73, 3331, 1994. 

\bibitem{SchoutensES-CFT} 
K. Schoutens, Phys. Rev. Lett. {\bf 79}, 2608 (1997)

\bibitem{Verlinde} E. Verlinde,
``A note on braid statistics and the non-Abelian Aharonov-Bohm effect'',
in {\it Modern Quantum Field Theory}
(World Scientific, Singapore, 1991). 

\bibitem{nonabelian_braiding} Goldin, R. Menikoff and D. H. Sharp, \journal 
Phys. Rev. Lett., 54, 603, 1985.

\bibitem{GLCSnonAbelian} E. Fradkin, C. Nayak, and K. Schoutens
``Landau-Ginzburg theories for non-Abelian quantum Hall states'', 
cond-mat/9811005 

\bibitem{AbelianGLCS}
S. C. Zhang, Int. J. Mod. Phys. {\bf B6}, 25 (1992).

\bibitem{Desbois} J. Desbois, C. Furtlehner, and S. Ouvry, Nuclear Physics B
	     {\bf 453 } [FS] 759 (1995).


\bibitem{Hagen}  
C. R. Hagen,  \journal  Phys. Rev. Lett., 76, 4086, 1996.

\bibitem{LeePRL} 
T. Lee,   \journal  Phys. Rev. Lett., 74, 4967, 1995; 
T. Lee and P. Oh, \journal Ann. Phys., 235, 413, 1994; see also
H.-K. Lo, \journal Phys. Rev. D, 48, 4999, 1993.

\bibitem{KZ}
V. G. Knizhnik and A. B. Zamolodchikov, Nucl. Phys. {\bf B 247}, 83 (1984).


\bibitem{JaPi} 
R. Jackiw and Y. S. Pi, \journal Phys. Rev. Lett., 64, 2969, 1990;
\journal Phys. Rev. D, 42, 3500, 1990; 
\journal Phys. Rev. Lett., 67, 415, 1991;
\journal Phys. Rev. D, 44, 2524, 1991.

\bibitem{Gir} Girvin et al. \journal Phys. Rev. Lett, 65, 1671, 1990.

\bibitem{JoCa} 
M. D. Johnson and G. S. Canright, Phys. Rev. {\bf B41}, 6870, 1990. 

\bibitem{BeLo} 
O. Bergamn and G. Lozano, \journal Ann. Phys., 229, 416, 1994; 
D. Z. Freedman, G. Lozano, and N. Rius, \journal Phys.Rev. D, 49, 1054, 1994.  

\bibitem{Stat} 
A. Dasni\`eres de Veigy and S. Ouvry, Nucl. Phys. {\bf B388} [FS], 715 (1992).
M. A. Valle Basagoiti, \journal Phys. Lett. B, 306, 307, 1993; 
R. Emparan and M. A. Basagoiti, \journal Mod. Phys. Lett A, 8, 3291, 1993.

\bibitem{Self} 
C. Manuel and R. Tarrach, \journal Phys. Lett. B, 268, 222, 1991;
\journal  Phys. Lett. B, 328, 113, 1994;
G. Amelino Camelia, \journal Phys. Lett. B, 299, 83, 1993; 
 {\bf 326}, 282, 1994; 
\journal Phys. Rev. D, 51, 2000, 1995; 
S. J. Kim and C. Lee, \journal Phys. Rev. D, 55, 2227, 1997; 
P. Giacconi, F. Maltoni, and R. Soldati, hep-th 9706198

\bibitem{Eza} 
Z. F. Ezawa, M. Hotta, A. Iwazaki, \journal Phys. Rev. D, 44, 452, 1991.

\bibitem{regularization}  
A. Comtet, Y. Georgelin and S. Ouvry, J. Phys. {\bf A}: Math. Gen. {\bf 22}, 
3917 (1989);
K. Olaussen, ``On the harmonic oscillator regularization of partition
function'', Trondheim Univ. preprint No.~13 (1992), 
cond-mat/9207005.

\bibitem{LIII} See, e.g.  L. D. Landau and E.M. Lifschitz, 
{\it Quantum Mechanics: Non-Relativistic} (Pergamon, 1977).


\bibitem{schur} See e.g.,  
S. Chaturvedi, \journal Phys. Rev. E, 54, 1378, 1996.  

\bibitem{PolychronakosNPB} 
A. P. Polychronakos,   \journal Nucl. Phys. B,  474, 529, 1996.

\bibitem{Schoutens-pfaffian} K. Schoutens, \journal Phys. Rev. Lett., 81,
1929, 1998. 

\bibitem{SIunpublished} S. B. Isakov, in preparation.

\bibitem{Poly} 
A. P. Polychronakos, private communication.

\bibitem{Kaplan} See, e.g. 
I. G. Kaplan, {\it Symmetry in Many-Electron Systems} (Academic Press, 1975).

\end{thebibliography}
\end{document}